\def\year{2022}\relax
\documentclass[letterpaper]{article} % DO NOT CHANGE THIS
\usepackage{aaai22}  % DO NOT CHANGE THIS
\usepackage{times}  % DO NOT CHANGE THIS
\usepackage{helvet}  % DO NOT CHANGE THIS
\usepackage{courier}  % DO NOT CHANGE THIS
\usepackage[hyphens]{url}  % DO NOT CHANGE THIS
\usepackage{graphicx} % DO NOT CHANGE THIS
\usepackage{natbib}  % DO NOT CHANGE THIS AND DO NOT ADD ANY OPTIONS TO IT
\usepackage{caption} % DO NOT CHANGE THIS AND DO NOT ADD ANY OPTIONS TO IT
\DeclareCaptionStyle{ruled}{labelfont=normalfont,labelsep=colon,strut=off} % DO NOT CHANGE THIS
\usepackage{amsmath}
\usepackage{amsfonts}
\usepackage{amssymb}
\usepackage[capitalise,noabbrev]{cleveref}
\usepackage{xcolor}
\usepackage{colortbl}
\usepackage{chngcntr}
\usepackage{multirow}
\usepackage{placeins}

\usepackage{newfloat}
\usepackage{listings}
\title{``If it didn't happen, why would I change my decision?'': How Judges Respond to Counterfactual Explanations for the Public Safety Assessment}
\author {
    Yaniv Yacoby,\textsuperscript{\rm 1}
    Ben Green,\textsuperscript{\rm 2}
    Christopher L. Griffin, Jr.,\textsuperscript{\rm 3}
    Finale Doshi-Velez\textsuperscript{\rm 1}
}
\affiliations {
    \textsuperscript{\rm 1} Harvard University \\
    \textsuperscript{\rm 2} University of Michigan\\
    \textsuperscript{\rm 3} James E. Rogers College of Law, University of Arizona\\
    yanivyacoby@g.harvard.edu, bzgreen@umich.edu, chrisgriffin@arizona.edu, finale@seas.harvard.edu
}

\begin{document}

\maketitle

\begin{abstract}
Many researchers and policymakers have expressed excitement about algorithmic explanations enabling more fair and responsible decision-making.
However, recent experimental studies have found that explanations do not always improve human use of algorithmic advice. 
In this study, we shed light on how people interpret and respond to counterfactual explanations (CFEs)---explanations that show how a model's output would change with marginal changes to its input(s)---in the context of pretrial risk assessment instruments (PRAIs).
We ran think-aloud trials with eight sitting U.S. state court judges, providing them with recommendations from a PRAI that includes CFEs.  
We found that the CFEs did not alter the judges' decisions.
At first, judges misinterpreted the counterfactuals as real---rather than hypothetical---changes to defendants. 
Once judges understood what the counterfactuals meant, they ignored them, stating their role is only to make decisions regarding the actual defendant in question. 
The judges also expressed a mix of reasons for ignoring or following the advice of the PRAI without CFEs. 
These results add to the literature detailing the unexpected ways in which people respond to algorithms and explanations.
They also highlight new challenges associated with improving human-algorithm collaborations through explanations.
\end{abstract}

\section{Introduction}
Many high-stakes decisions are now made by people with the aid of algorithmic advice, including in the labor market and the criminal justice system. 
In order to promote more informed and responsible human uses of algorithms, many researchers, policymakers, and engineers have expressed interest in explainable AI (XAI).  
The goal of XAI systems is for algorithms to provide reasons behind the recommendations they generate.  
In theory, such systems could help human decision-makers identify errors, correct for biases, and synthesize the system's reasoning with their own (e.g.,~\citet{ribeiro2016should, Lundberg2017, adler2018auditing, Bach2015, guidotti2018survey}).  
This has motivated regulatory bodies to recommend explanations for high-stakes AI systems (e.g.,~\citet{engstrom2020government,expertgroup2019,uk2020}).  

Alongside this excitement about XAI, however, recent work has demonstrated that these benefits are not always realized in practice. 
User studies in multiple contexts have found that explanations did not help people evaluate the quality of algorithmic advice or incorporate that advice into decisions (e.g., \citet{lai2019human, bansal2021does, jacobs2021machine, Green2019Principles}). 
Thus, it is not yet clear whether and under what conditions explanations improve human-algorithm decision-making.  
Moreover, most prior user studies have been limited to laypeople. 
It is particularly important to understand how practitioners interact with algorithmic explanations.

In this study, we investigated how sitting U.S. state court judges interact with a type of explanation known as ``counterfactual explanations'' (CFEs).
CFEs provide a human decision-maker with information about how a model's output changes based on variations in the input(s) (e.g., \citet{martens2014explaining, wachter2018counterfactual, Goyal2019, verma2020counterfactual, stepin2021survey}). 
In theory, this information could provide human decision-makers with a better understanding of how sensitive---or robust---the model is to marginal changes to its input(s).  
This insight could then help decision-makers determine whether and how to use the model's predictions.

We focused on CFEs for algorithmic pretrial risk assessment instruments (PRAIs).  
PRAIs take information about a defendant (e.g., age, prior failures to appear) as inputs and calculate one or more risk scores (e.g., the defendant's likelihood of being arrested if released pretrial) to guide pretrial detention decisions (e.g., whether to release or detain the defendant). 
Although PRAIs are often touted as mitigating human biases, the underlying models themselves have been shown to perpetuate systematic biases and reproduce structural inequities (e.g.,~\citet{angwin2016machine, Green2020False, koepke2018danger, barabas2018interventions}).  
Given these concerns, PRAIs present a salient application in which policymakers and scholars focus on XAI \cite{green2022flaws}.

We investigated whether CFEs alter how judges understand and respond to a PRAI.
For example, consider a pretrial defendant who was labeled ``high risk'' by a PRAI that provides CFEs. 
A  CFE might reveal that this classification was a function of a prior felony conviction (a disproportionate outcome for non-white defendants): had the defendant not been convicted, they would have been deemed ``low risk.''  
Would such information---instead of just the risk score---encourage a judge to inquire about the charge behind the felony conviction? If so, might they lower their risk estimate if the charge were for a non-violent crime? 
Alternatively, suppose the CFE revealed that marginal changes to a defendant's criminal history significantly would change the defendant's risk classification; sometimes up, sometimes down.  
Would this information encourage a judge to deem the risk assessment less reliable overall?
Furthermore, would any of these reactions be swayed by (perhaps implicit) bias about the defendant's race, only exacerbating existing systemic bias?

To explore how judges respond to CFEs, we conducted qualitative think-aloud studies with eight sitting judges in two U.S. states that use a PRAI called the Public Safety Assessment (PSA).
We presented judges with hypothetical pretrial cases, along with the PSA report and CFEs, and prompted them to share their reasoning in real time. This provided us with detailed knowledge about how judges reason about PRAIs and CFEs.

The judges responded to the CFEs in both unexpected---and unexpectedly consistent---ways. 
Initially, the judges were unsure about how to interpret the CFEs, assuming that they presented factual changes to the defendants' profiles. 
Once they understood that the explanations represented hypothetical changes to the defendants' profiles, they ignored them, claiming the need to focus on the defendant at hand.  They did not consider what the explanations signaled about the model's sensitivity. 
This behavior persisted even after we provided explicit coaching about what information the explanations contained and how it could be used.

We hypothesize that judges had difficulty understanding the CFEs not because they lack training in statistics or AI, but because of their legal training. 
Judges are trained to think counterfactually---but only about the facts of the case before them, not facts about the defendant. When considering whether a defendant caused a victim's injury, for example, a judge might naturally ask whether the injury would have occurred without the defendant's actions. A judge will not question whether the victim might have been uninjured had the defendant been a different person who had different interactions with the criminal justice system before encountering the victim.
Thus, the counterfactual reasoning judges employ focuses on particular actions of the \emph{defendant}, whereas CFEs provide insight about the reasoning of the \emph{model}.

Our findings contribute to a growing body of literature suggesting that people react to explanations in often unexpected ways. 
While our study sample was relatively small, the use of practitioners as participants and the consistency of the judges' reactions provide new insights into the challenges associated with achieving any benefits from CFEs.

\section{Related Work} \label{sec:related-work}

\paragraph{Pretrial Risk Assessment.}
PRAIs have gained traction as a tool for criminal justice reform efforts, but they suffer from a variety of flaws.  
PRAIs are typically hailed for replacing biased human judgments with ``objective'' algorithmic judgments (e.g., \citet{arnold2019faqs, nj2017report}), enabling more consistent and fair decision-making. 
However, PRAIs depend on data that reflect human biases and systemic inequities, thus perpetuating these biases and inequities (e.g.,~\citet{propublica2016, barabas2018interventions, Green2020False, koepke2018danger}). 
Furthermore, rather than eliminate human discretion, PRAIs shift discretion to different actors and decision points~\cite{Green2020False}.
Broadly, because PRAIs legitimize policies associated with mass incarceration, critics have challenged their ability to promote decarceral criminal justice reform \cite{Green2020False, koepke2018danger, barabas2018interventions}. 

The interaction between PRAIs and human decision-makers poses another significant challenge for criminal justice reform. 
Experimental research demonstrates that laypeople respond to PRAIs in a racially biased manner, being more susceptible to increasing risk estimates following a risk assessment's outputs when evaluating Black defendants relative to white defendants \cite{Green2019Disparate, Green2019Principles}.  
Similar behaviors seem to manifest in practice. 
Judges across U.S. jurisdictions frequently override the PRAI's outputs that recommend releasing defendants, leading to much higher pretrial detention rates than expected when risk assessments are used \cite{hrw2017, sheriff2016, steinhart2006practice, stevenson2018assessing, stevenson2021algorithmic}. 
Furthermore, counter to expectations, the use of PRAIs has increased racial disparities because judges respond to recommendations in more punitive ways when evaluating Black as opposed to white defendants \citet{albright2019judge, cowgill2018impact, stevenson2021algorithmic}.  

An important limitation of prior work on human interactions with PRAIs is a lack of knowledge about the thought processes that criminal justice practitioners follow when incorporating PRAI recommendations into their decisions.
Earlier studies have primarily relied on lab experiments with laypeople (e.g., \citet{Green2019Disparate,Green2019Principles, grgic2019human}) or empirical analyses about judicial decisions using observational data (e.g., \citet{albright2019judge, stevenson2018assessing}). 
Thus, with a few exceptions \citep{brayne2020technologies, hannah2009negotiated}, we have little understanding of the thought processes that criminal justice practitioners follow when incorporating PRAI advice into their decisions. 
Our study contributes to research on the implementation of PRAIs by leveraging detailed, real-time descriptions of how judges reason about these models.

\paragraph{Explanations and Their Challenges.}
Previous work studying how people react to algorithmic explanations demonstrates that different ways of presenting explanations can alter how people perceive the fairness of those decisions (e.g., \citet{binns2018reducing, dodge2019explaining}).
Furthermore, recent work shows that explanations do not always improve people's ability to effectively use algorithmic predictions (e.g.,~\citet{bansal2021does, Green2019Principles}).  
Explanations can increase human trust in the algorithm's recommendations, even when these recommendations are incorrect (e.g., ~\citet{bansal2021does,jacobs2021machine}) or when the explanations do not accurately represent the algorithm's inner workings (e.g.,~\citet{lai2019human}).
Our work adds to this literature by examining how practitioners respond to explanations in a specific, high-stakes (albeit artificial) setting: judges making pretrial release decisions.  

\paragraph{Counterfactual Explanations (CFEs).}
A CFE is a statement of the form, ``if input $x$ would have been altered by a small change (now called $x'$), a decision making system would output $y'$ instead of $y$''~\cite{wachter2018counterfactual}. In this statement, $x, y$ are the ``factual'' input and output and $x', y'$ are the ``counterfactual'' input and output. 
Formally, CFEs are the solutions to an optimization problem of the form,
\begin{equation}
    \mathrm{argmin}_{x'} d(x, x') \quad \text{subject to} \quad f(x') = y',
    \label{eq:counterfactual}
\end{equation}
where $d(\cdot, \cdot)$ is a distance metric, $f(x')$ is the output of the decision making system, and $y'$ is the desired outcome \cite{verma2020counterfactual}. 
Looking at the set of inputs for which the decision making system outputs $y'$ (i.e. all $x'$ for which $f(x') = y'$), the above optimization problem selects the $x'$ that is closest to $x$ (according to distance metric $d(\cdot, \cdot)$).
By framing the problem in this way, we find the ``minimal'' change from $x$ to $x'$ that will change the outcome from $y$ to $y'$.
Counterfactuals generated in this way follow guidelines from psychology (e.g., ~\citet{miller2017explainable,keane2021counterfactual}) to be maximally intuitive by ensuring they are both ``sparse'' (requiring few features to change) and ``proximate'' (close to the original factual).

A growing body of literature (e.g., ~\citet{miller2017explainable, wachter2018counterfactual, byrne2019counterfactuals, mueller2019explanation}) calls for the use of CFEs due to the importance of counterfactual reasoning in human thought, as historically demonstrated in fields such as psychology (e.g., ~\citet{kahneman1981simulation, roese1997counterfactual, mccloy2000counterfactual, hilton2005course, byrne2007rational, byrne2016counterfactual}) and philosophy (e.g., ~\citet{woodward2005making, lewis2013counterfactuals}), and because CFEs potentially comply with emerging regulations governing the use of AI~\cite{wachter2018counterfactual}.

Despite this growing interest in CFEs, we do not yet know whether, when, and how CFEs enable people to make more accurate and fair decisions given algorithmic advice. 
Notably, two recent surveys of CFEs call out a significant lack of user studies testing how people respond to CFEs \cite{verma2020counterfactual, keane2021counterfactual}.
Our study adds to the literature by providing evidence about how practitioners interpret and respond to CFEs when reasoning about high-stakes decisions in a laboratory setting.

\section{The Public Safety Assessment (PSA)}

In this section, we provide background on current PRAIs and how they are integrated into pretrial release decision-making.

\paragraph{The Public Safety Assessment (PSA) Algorithm.}
The PSA is a PRAI developed by Arnold Ventures to ``provide judges with objective and consistent data to make informed decisions'' about pretrial defendants \cite{arnold2019principles}.  
The PSA takes static information about the defendant's age, the pending charge, and seven measures of prior criminal justice experiences as inputs. It assigns these risk factors an initial set of integer weights, which are then converted into two risk scores---the New Criminal Activity (NCA) and Failure to Appear (FTA) scores. These risk scores range from 1 to 6, with larger values indicating higher risk. 
The NCA score reflects the risk that the defendant will be arrested for a new charge if they are released, and the FTA score reflects the risk for the defendant’s failure to appear to a subsequent hearing if they are released.
Lastly, the new violent criminal activity (NVCA) flag provides a binary indicator that the risk of arrest for a new violent charge is high.
These scores are presented to judges before or during pretrial hearings, at which judges must decide whether to release (with or without conditions, including money bail) or detain a defendant before trial.

\definecolor{ROR}{RGB}{101, 130, 49}
\definecolor{PML1}{RGB}{166, 207, 91}
\definecolor{PML2}{RGB}{252, 253, 113}
\definecolor{PML3}{RGB}{242, 206, 73}
\definecolor{PML3P}{RGB}{224, 148, 76}
\definecolor{NoRelease}{RGB}{172, 92, 82}

\begin{figure}[t]

    \scriptsize
    \centering
    \setlength{\tabcolsep}{0.5em}
    
    \begin{tabular}{cc||cccccc|}
      & & \multicolumn{6}{c}{\textbf{NCA}} \\
      & & \textbf{1} & \textbf{2} & \textbf{3} & \textbf{4} & \textbf{5} & \textbf{6} \\ \hline \hline
      \multirow{6}{*}{\rotatebox[origin=c]{90}{\textbf{FTA}}} & \textbf{1} & \cellcolor{ROR} ROR & \cellcolor{ROR} ROR & & & & \\ 
      & \textbf{2} & \cellcolor{ROR} ROR & \cellcolor{ROR} ROR & \cellcolor{PML1} PML1 & \cellcolor{PML2} PML2 & \cellcolor{PML3} PML3 & \\ 
      & \textbf{3} &  & \cellcolor{PML1} PML1 & \cellcolor{PML1} PML1 & \cellcolor{PML2} PML2 & \cellcolor{PML3} PML3 & \cellcolor{NoRelease} No Release \\
      & \textbf{4} &  & \cellcolor{PML1} PML1 & \cellcolor{PML1} PML1 & \cellcolor{PML2} PML2 & \cellcolor{PML3} PML3 & \cellcolor{NoRelease} No Release \\
      & \textbf{5} &  & \cellcolor{PML2} PML2 & \cellcolor{PML2} PML2 & \cellcolor{PML3} PML3 & \cellcolor{PML3P} PML3+EM/HD & \cellcolor{NoRelease} No Release \\
      & \textbf{6} &  &  &  & \cellcolor{NoRelease} No Release & \cellcolor{NoRelease} No Release & \cellcolor{NoRelease} No Release \\
      \hline
    \end{tabular}
    
    \caption{\textbf{Reproduction of DMF Matrix from the~\citet{dmf_nj}.} This matrix maps the risk scores generated by the PSA---the FTA and NCA scores---to a color-coded release recommendation ranging from least restrictive (green) to most restrictive (red). We note that the matrix is also accompanied by additional instructions depending on the charges filed. For example, for a murder or felony-murder charge, the instructions will recommend ``no release'' regardless of the PSA's outputs,  (\url{https://www.njcourts.gov/courts/assets/criminal/decmakframwork.pdf}). 
    Key: ROR = release on their own recognizance; PML = Pretrial Monitoring Level; EM = Electronic Monitoring; HD = Home Detention.}
    \label{fig:example-dmf}
\end{figure}

\paragraph{The Decision-Making Framework (DMF).}
The NCA and FTA risk scores generated by the PSA (and the NVCA flag) are inputs into the Decision-Making Framework (DMF). 
In short, the DMF converts the three indicators into a recommendation for release conditions, if any. 
The central component of the DMF is the DMF Matrix, which combines the NCA and FTA scores to yield recommended release conditions (see \cref{fig:example-dmf}).
Recommendations can diverge from a jurisdiction's DMF Matrix based on the type of charge and the NVCA flag. 
Local stakeholders (e.g., the court, public defenders, prosecutors) collectively choose the recommendations corresponding to combinations of NCA and FTA scores. 
These choices usually reflect policy goals and values in the jurisdiction, statutory requirements, and available resources (e.g., pretrial services). 
Thus, while the PSA scores in all jurisdictions rely on the same data and are the product of the same scoring algorithm, the DMF varies across jurisdictions. 
Finally, decisions about releasing or detaining defendants (and under what conditions) always remain subject to judicial discretion.  
In this work, when referring to the ``model,'' we mean the combined PSA and DMF system (PSA-DMF).

\section{Methodology}
We designed a survey to assess how judges respond to PSA-DMF recommendations and to CFEs of those recommendations. 
In the main body of the survey, we asked participating judges to make release decisions for synthetic defendants.
Judges were first asked to make these decisions based on factual information about defendants (e.g., age, criminal history) and the PSA-DMF reports. 
Judges were then presented with CFEs (as additional information) and had the opportunity to update or stand by their initial decision. 

We will first describe how we generated defendant data, DMF recommendations, and CFEs. We will then describe the participants we recruited and the experimental protocols. 

\subsection{Experiment Setup}
\begin{figure*}[t]
    \centering
    \includegraphics[width=0.71\textwidth]{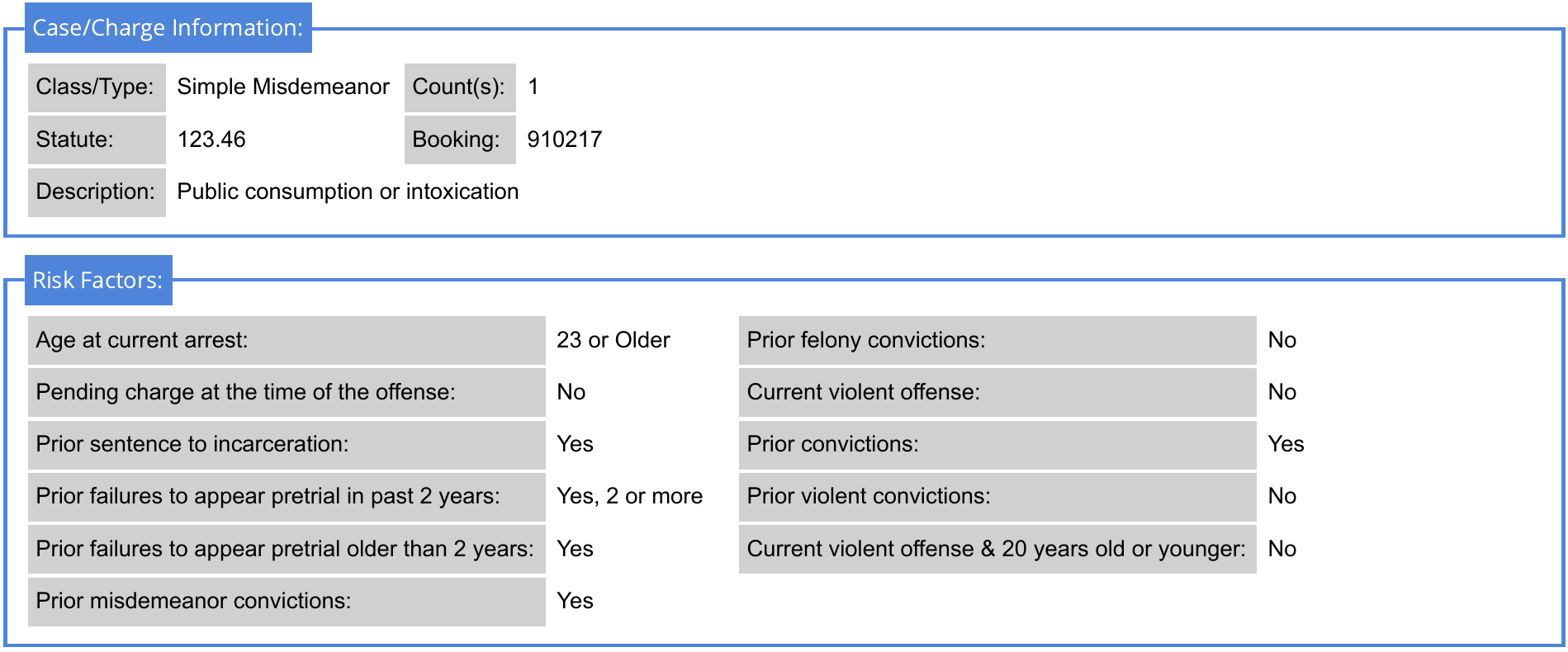} 
    \caption{\textbf{Example case/charge information and risk factors from our survey.}}
    \label{fig:charge-info-and-risk-factors}
\end{figure*}

\paragraph{Defendant Data.}
We created a set of realistic cases based on a sample of 500 de-identified cases from a county in Iowa that had used the PSA-DMF, consisting of actual defendants' statutory charges, risk factor values, and PSA-DMF recommendations.
To ensure judges were not presented with a homogeneous set of cases, we selected a subset of cases with a diverse set of common charges, and for which the CFEs included a variety of hypothetical changes leading to more or less restrictive recommendations (or both).
Lastly, we synthesized the demographic information, including the date of birth, date of arrest, race, and name (ensuring the name was likely perceived to correspond to the defendant's race, using the top first names identified by~\citet{gaddis2017} and top last names identified by~\citet{word2008}).
See \cref{sec:defendant-data} for more details, and see \cref{fig:charge-info-and-risk-factors} for an example defendant.

\begin{figure*}[t]
    \centering
    \includegraphics[width=0.71\textwidth]{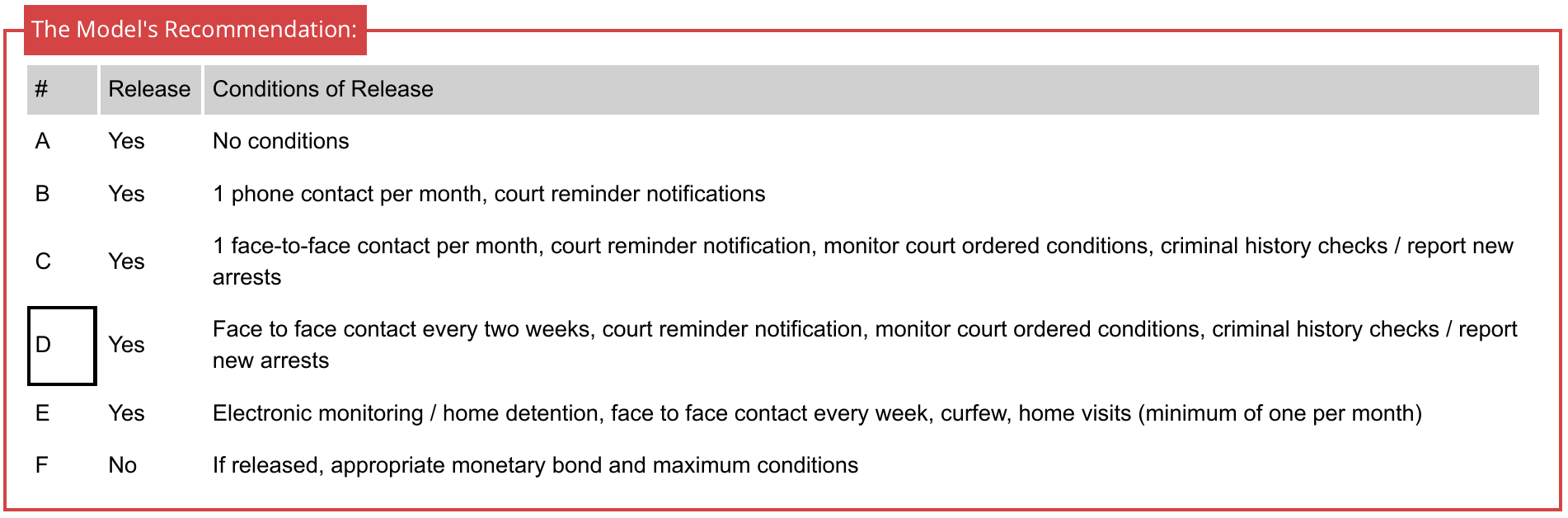} 
    \caption{\textbf{Example DMF recommendations from our survey.} In this example, the model (PSA-DMF) recommends option `D', which is for the defendant to be released, but with face-to-face contact every two weeks, etc.}
    \label{fig:release-recommendation}
\end{figure*}

\begin{figure*}[t]
    \centering
    \includegraphics[width=0.73\textwidth]{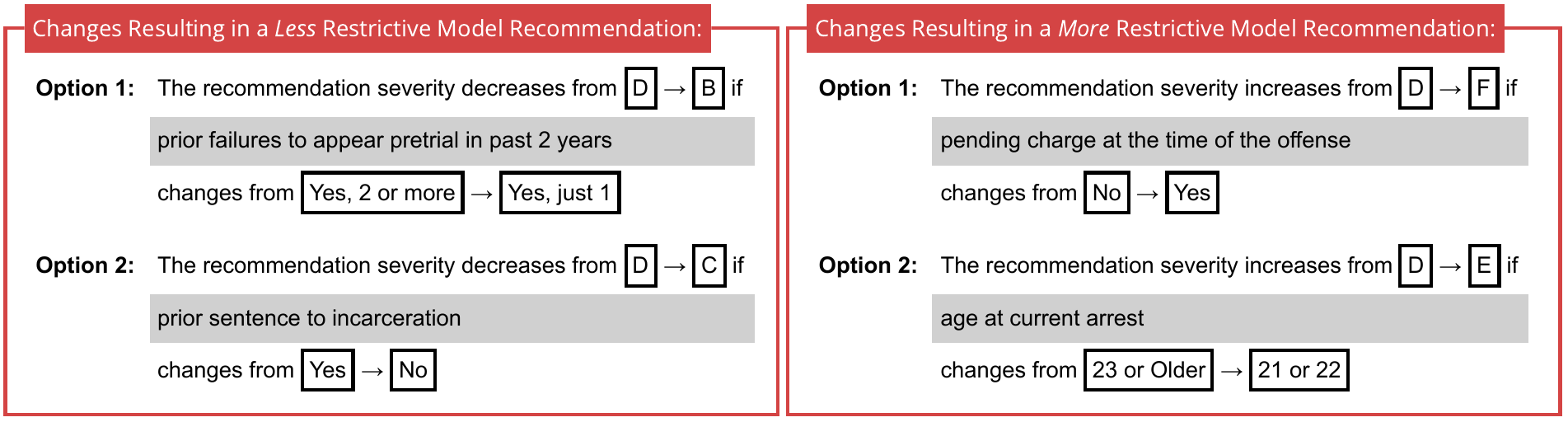} 
    \caption{\textbf{Example CFE from our survey.} The two red boxes present single changes to the defendant's profile that would decrease and increase, respectively, how restrictive the release recommendation is. Inside each of the red boxes are several options, each of which describes the single change to the defendant's profile that would change the model's recommendation.}
    \label{fig:counterfactual-explanation}
\end{figure*}

\paragraph{Generating DMF Recommendations.}
We used a DMF matrix similar to the one in \cref{fig:example-dmf}.
Since we generated counterfactuals of the combined PSA-DMF relative to the PSA's \emph{inputs} (see below), we chose not to include a full visualization of the DMF matrix, which illustrates how the DMF recommendations change relative to the PSA's \emph{outputs}.
Instead of using a matrix of colors and acronyms for the release options (e.g., ROR, PML1, etc.), we assigned the different release options letters A through F (from least to most restrictive) and presented them as a list---see \cref{fig:release-recommendation}.

\paragraph{Generating Counterfactuals.}
Following the guidelines from \cref{sec:related-work}, we designed the counterfactuals to be intuitive by ensuring they were both sparse and proximate.
For each defendant in the cohort, we iterated over all possible risk factor combinations (with only nine factors, one can easily iterate through all of them), keeping only risk factor combinations that were (a) consistent with one another, (b) a single ``edit'' different from the original defendant, and (c) for which the model's final recommendation differs from the original recommendation.
Example counterfactuals are visualized in \cref{fig:counterfactual-explanation};
we chose this representation for its concision relative to other types of visualizations (e.g., heatmaps). 

\subsection{Think-Aloud Protocols}
\paragraph{Participants.} 
We recruited eight judges to participate in our study---four judges from a Mountain region state for Round 1 and four judges from a Southwest region state for Round 2---via direct solicitation and snowball sampling. All participants were judges in districts that use the Arnold Ventures PSA-DMF.  
Thus, all participants already had regular experience with the model, to which our study added the CFEs. 
Participants had between $3$ and $26.5$ years of experience on the bench (mean of $16.4$ years).  

\paragraph{Study Overview.}
The think-aloud protocol for both rounds of the survey are included in \cref{sec:round1-survey,sec:round2-survey}; 
screenshots of the survey are included in \cref{sec:round2-survey-screenshots}, along with example cases in \cref{sec:additional-cases}. 
After obtaining consent, participants received a short tutorial on our adapted PSA-DMF report (\cref{fig:r2-instructions1,fig:r2-instructions2}), including what CFEs are and how to read them. 
Next, to ensure participants understood the information presented in the CFEs, we asked basic comprehension questions about the CFE (e.g., ``If the given defendant had one less prior conviction, what would the model recommend?'')---see \cref{sec:correctness-comprehension} for details. 

After the consent process, instructions, and comprehension questions, we moved to the main body of the survey (\cref{fig:r2-question1pre,fig:r2-question1post,fig:r2-question2pre,fig:r2-question2post}).
We presented each synthetic defendant's risk factors, demographic information, and the model's release recommendation (without any explanation).
We then asked participants to narrate their thought processes while reading and interpreting the PSA-DMF report before arriving at their release decision. After formulating that release decision, we showed them the CFE and asked them whether they wanted to revise their original decision based on the new information (and if so, why). For those who were hesitant to think aloud or who stopped sharing in the middle of the survey, we prompted them to continue by asking questions such as, ``What led you to choose that option?''

For visual distinction, we placed all model outputs in red boxes, i.e., the FTA and NCA scores and DMF recommendations, as well as the CFEs (\cref{fig:release-recommendation,fig:counterfactual-explanation}). We placed all remaining information, i.e., the inputs to the model (risk factors), charge and demographic information, in blue boxes (\cref{fig:charge-info-and-risk-factors}). 
We note the colors here because some judges referred to the boxes by color. 

We conducted all interviews via Zoom and limited them to about one hour to respect the judges' time.  Judges completed an average of $4.4$ of the cases in that time, and, with their permission, we recorded every interview.
We had each interview professionally transcribed.
Two members of the research team then reviewed these transcripts to characterize the views expressed by judges during the interviews.
We present each finding with quoted language from multiple participants in order to demonstrate the common threads across responses. 
This work was approved by the [Anonymous] Institutional Review Board. 

\paragraph{Differences between Rounds 1 and 2.}
When conducting interviews with the first four judges (Round 1), we found surprising yet consistent results. As we describe in more detail in \cref{sec:findings-explanation}, even with the tutorial and comprehension questions, the judges interpreted CFEs as new \emph{factual} information: they thought that the actual defendant's profile had been updated. Once they understood that the CFEs contained hypothetical scenarios, the judges still largely ignored them. They believed it inappropriate to consider hypothetical information about a defendant when making decisions about release---even though the hypotheticals were about the PSA-DMF outputs. In other words, the judges mistakenly confused the CFEs as indicative of a new defendant before them, as opposed to new information about the functioning of the PRAI using the same defendant.

These findings prompted us to adjust our protocol. In Round 1, we deliberately excluded specific training on how the judges might use the counterfactuals to adjust their decisions, because we did not want to influence their behavior. However, our initial results prompted us to ask whether judges would respond differently if we provided more training on what the counterfactuals meant and how judges might use them.  Although doing so might have primed the judges to use the explanation in certain ways, it would also help us explore whether the results in Round 1 were due to a lack of understanding of our instructions or actual behavior.

For Round 2 (detailed in \cref{sec:round2-survey}), we therefore expanded our training to provide more information about the CFEs. We created a presentation to describe what a CFE was and concrete ways in which it could be used (e.g., if the defendant's age is 22 or 23 years, the judge can use the CFE to ``adjust'' the final recommendation, depending on whether they believe that age is an important factor). In the presentation, we additionally asked judges to explain to us what the CFE was to ensure they understood it. We hoped that having a live, interactive presentation at the start of the survey would ensure that participants paid attention (rather than potentially skimmed the tutorial) and felt comfortable asking clarifying questions (as opposed to feeling self-conscious about their reading and comprehension speed). Indeed, the judges felt more comfortable asking questions; when asked to explain the counterfactuals to us, nearly all of them did so well.

\section{Results} 
Across all eight think-aloud participants (four in each of the two rounds), we observed notable consistency in how the judges used the PSA-DMF and CFEs. 
We therefore present the results from both rounds jointly.
As a reminder, when the judges refer to the ``red boxes,'' they are referring to information based on the output of the PSA-DMF (or model), and when they refer to the ``blue boxes,'' they refer to all other information, including the inputs to the PSA-DMF. 

\subsection{How Judges Use the PSA-DMF} \label{sec:findings-psa}
Because we first asked judges to evaluate each defendant with the PSA-DMF report but without the CFEs, our think-aloud interviews provided insights into how judges reason when making PRAI-assisted pretrial release decisions. 
Judges' responses to the model's information revealed mixed attitudes regarding the utility and quality of the PSA-DMF.

\paragraph{Judges focused on the specific attributes of defendants, rather than the risk scores.}
When thinking aloud about release decisions, judges focused first on the charge class, i.e., whether the charge was a felony or misdemeanor, and the violence flag, noting that the charge class helped them determine whether the defendant is a risk to society. Next, they scanned the risk factors only to check if they were unusual or inconsistent with other information; otherwise, they ignored them and continued to the DMF's release recommendation.

\begin{quote}
``I kind of pretty much disregarded the risk scores. [...] I gave the risk assessment a glance, but I was more interested in the detail of the risk factors. [...] Many judges want to see the details of the thing as opposed to focusing on in what we're calling here the risk assessment itself. It's of interest, but the blue box is more interesting to me.'' -- P1 
\end{quote}
\begin{quote}
 ``I mostly relied on the blue boxes, less on the red boxes.'' -- P2
\end{quote}
\begin{quote}
``The risk factor box is the one that's frankly the most important to me. And, what I'm looking at and what is very important to me when I look at this is typically pending charges at the time of the offense, failures to appear, the two failure to appear boxes, obviously the violent offense boxes. These are the ones that are most important to me. Secondarily to that is probably the sort of where you land on the matrix [the FTA and NCA scores]. '' -- P4 
\end{quote}
\begin{quote}
``I'm looking at the kind of charge it is, I'm looking at the PSA. I'm looking at whether there's a victim, whether it's a violent offense, whether somebody died. If it's a violent offense, the more likely it is that they're going to be held. If it's a sexual offense, and I see a lot of PSAs where they recommend release, personally they're probably going to be held [on a high bond] despite the recommendation.'' -- P5
\end{quote}
\begin{quote}
``To me, a big part of my decision making is based on the facts that are alleged in the case.'' -- P7
\end{quote}

\paragraph{Some judges expressed concerns about the accuracy of the risk scores associated with the PSA model.} In particular, judges expressed concerns about specific cases, for example those involving domestic violence charges.
\begin{quote}
    ``I never look at these [PSA recommendations], honestly, when I'm making decisions in court. [...] I know that the [colleagues] I've talked to, we don't spend a lot of time looking at the PSA. [...] Sometimes the recommendations strike us as very off, given what other information we have. [...] It's not something I don't think any of us spend the majority of our time on, when we're making these decisions. [...] I feel like the PSA is totally, totally not valid for domestic violence cases.'' -- P7
\end{quote}
\begin{quote}
    ``[Given a defendant with no previous criminal history, but with age $\leq 20$:] This is actually one of the things I disagree with on PSA because it tells me new criminal activity score. And he's got him for a[n NCA of] two, but he's only got the one charge. So how would you have a one [charge], if you're saying it's new criminal activity based on the actual charge you're seeing the guy on.'' -- P8
\end{quote}
Here, the judge is concerned that a defendant with no prior criminal history is given an NCA of 2 (likely because younger defendants are empirically at higher risk of being re-arrested after release).

\paragraph{Judges also described reviewing and deferring to the PSA-DMF recommendations.}
Notably, although both P2 and P4 had stated that they primarily focus on the case details, they also described regularly considering and deferring to the recommendations.
P2 described deferring to the PSA-DMF model suggestion for cases at the extremes of the risk score values.
\begin{quote}
``The more polar the recommendation: if I'm on a one, then I'm going to cut to the one and I'll be good with the one. And the same with the six, I'll go, ``Yeah, the guy's got a ton of risk factors.'' Now, if I'm in the 3, 4, 5, then I'm going to start going back and looking at the individual risk factors.'' -- P2 
\end{quote}
This quote describes behavior that is consistent with the findings of a prior experimental study regarding how laypeople respond to PRAIs \cite{green2021alter}.

P4 described following the DMF recommendation in the interest of consistency.
\begin{quote}
``I would tend to default where there's not a huge difference [between my and the PSA's recommendations]. [...] For purposes of consistency, I would tend to default with what the matrix recommends.'' -- P4 
\end{quote}
Similar to the prior quotation, this comment suggests that judges selectively defer to the PSA-DMF, based on the observed risk score values and their own internal assessments.

Finally, one judge noted that they were explicitly asked to follow the DMF release recommendation during implementation training to facilitate a social science evaluation of the PSA-DMF's efficacy.
\begin{quote}
``So, when we deal with PSAs, we've been, because it's kind of a---for lack of a better term---a pilot program, they've asked us to go with the recommendations of the PSA. So when I'm making a decision, often even if I think that the recommendation [...] is a lot to ask someone to do [...] I've been asked to defer to the PSA. [...] All of the training that I've gone to kind of says ``less is more'' on a lot of these types of cases, but we've been asked to go with the PSA because they've been trying to assess how effective those are.'' -- P3 
\end{quote}
This passage conflicts with the claim by Arnold Ventures and policymakers that judges are not required/expected to follow DMF recommendations \cite{arnold2019faqs, nj2017report}---more detail in \cref{sec:discussion}.

In sum, judges neither completely disregarded nor completely followed the PSA-DMF recommendations. 
Instead, judges reasoned about the case at hand and compare their own assessment with the assessment provided by the model. 
Several judges who explicitly claimed to focus on the case details over the model's report generally also described considering the model’s recommendations when making final decisions. 
Because judges (claim to) use the model's recommendation differently depending on the particular details of each individual case, it is particularly important to understand whether and how model explanations alter this decision-making process. 

\subsection{How Judges Use Counterfactual Explanations} \label{sec:findings-explanation}
After judges made initial decisions, we presented them with CFEs and asked whether they would like to update their decisions.

\paragraph{The judges initially treated counterfactuals as factuals.}
When presented with the CFEs, judges almost always treated the counterfactuals as real changes to the defendant's profile, rather than as hypothetical changes, and updated their decisions based on this information.
\begin{quote}
``I was assuming in answering that, that I was supposed to assume that one of these risk factors had changed.'' -- P4
\end{quote}
\begin{quote}
[When viewing a counterfactual in which a defendant had one FTA instead of two:] ``So now I'm going down and I see that he's only got one failure to appear. Is that what I'm looking at now?'' -- P8
\end{quote}
As a result, judges were particularly bothered when the counterfactual included a pending charge, because they wanted to factor the alleged crime into their decision-making process. It was difficult for judges to reconcile that, because the CFE applies to the model, \emph{any} pending charge---whether serious or not---would alter the model's recommendation in the exact same way. This phenomenon follows the behavior described in \cref{sec:findings-psa}, where judges focused on the charge details when making pretrial release decisions. 
\begin{quote}
[Given a CFE in which an additional failure to appear leads to a `B' and an additional pending charge leads to a `C' from an `A':] ``It's more so the pending charge rather than the failure to appear. I mean it's a factor, but it's not as big of a factor as the pending charge.'' -- P5 
\end{quote}
\begin{quote}
[Looking at the CFE, consisting of 5 options leading to more restrictive recommendations:] ``If there is a pending charge at the time of the offense, that could certainly change [my decision]'' -- P6 
\end{quote}
\begin{quote}
``If he has a pending charge at the time of the offense, well, I would want to know what the actual pending charge was.'' -- P8
\end{quote}

\paragraph{Interpreting counterfactuals as factuals confused judges about how counterfactuals could generate changes in both directions.}
Given that judges interpreted counterfactuals as real changes to the profile of the defendant in question, they were particularly concerned by sets of counterfactuals involving changes to the defendant's profile that led to both more and less restrictive release recommendations. 
Of course, if these changes were factual, then they could not be simultaneously true.
\begin{quote}
``Ok I'm not clear---I'm looking at one set that says `less restrictive' and one that says `more restrictive', and I'm sorry for being dense here, but I don't have sufficient questions for me to answer each one of them under the `less' model and each one of them under the `more' model, and my answers would be different depending on whether the less restrictive recommendations are in place or the more restrictive recommendations are in place. So I'm not sure what to do---I'm at an impasse. [...] It appears to me that you're offering me two choices and one set of answers. [...] I don't see that I have a way to answer both the less restrictive and the more restrictive.'' -- P1
\end{quote}
\begin{quote}
``All right, now I'm a little confused by the instrument. [...] So if I got two red boxes, I got one going less and I got one going more, which release conditions... [...] and my box on the left, option one drops him to a two, [... but the question] doesn't tell me which one he is. Is he the less one or is he the more one?'' -- P2
\end{quote}
\begin{quote}
[After receiving the CFE:] ``Am I supposed to take into account both options 1 and 2? [...] Are these two taken collectively? Both option 1 and option 2 have changed now?'' -- P3
\end{quote}
\begin{quote}
``So when I'm answering this question, am I answering it for the red box on the left or the red box on the right?'' -- P8
\end{quote}

\paragraph{Once judges understood that counterfactuals were hypothetical, they generally ignored them.}
When judges seemed to be misinterpreting CFEs as factual, we emphasized that the counterfactuals were hypotheticals designed to illustrate the properties of the algorithmic PRAI, not an actual revision to the defendant's profile. 
At this point, the judges stopped considering the counterfactuals altogether. 
\begin{quote}
[After realizing that counterfactuals are not revisions to the defendant's profile:] ``I would still choose the same. No change.'' -- P3 
\end{quote}
\begin{quote}
``Oh! Oh! [...]. I was assuming in answering that, that I was supposed to assume that one of these risk factors had changed. [...] I don't know what I put in response to the other ones, but these should be identical. [referring to the questions before and after the explanation].'' -- P4 
\end{quote}
\begin{quote}
``No [I would not change my decision], because nothing has changed. It's a little awkward thought process. [...] What you are saying is: these didn't happen. If they didn't happen, [...] how will I change my recommendation or my decision? That's where I'm losing the logic. [...] If it didn't happen, why would I change my decision?'' -- P6 
\end{quote}
\begin{quote}
[After realizing the defendant's profile has not changed:] ``No, then if he's the same person he was to begin with, then I would still just release him [...] because my decision is mostly based on the charge itself and the lack of failures to appear or anything like that, so I think I would leave it the same.'' -- P7
\end{quote}
\begin{quote}
``If the person's different, that's definitely relevant to me. So if these things, hypothetically, if in fact these things are true, I might change my recommendation. [...] [After confirming that the CFEs are merely hypothetical:] Well, I'm going to do the same thing [as before viewing the counterfactuals]. I'd just even be more firmly convinced I was right.'' -- P8
\end{quote}

\paragraph{One judge described circumstances in which they could envision changing their responses based on counterfactuals.}  
This judge said that they might revisit their original decision, placing less trust in the PSA-DMF, if a hypothetical change to the defendant's profile caused a large change to the DMF recommendation (e.g., from release option B, ``1 phone contact per month, court reminder notifications'' to E, ``Electronic monitoring/home detention, etc.''):
\begin{quote}
``If that one factor is enough to move it that far, then that would cause me concern that the original recommendation is not restrictive enough.'' -- P4  
\end{quote}
\begin{quote}
[When asked: If the counterfactual showed a several step increase or decrease in the recommendation, would your decision change?] ``Yeah. I would probably change my original condition. [...] It would give me some pause about the underlying math, if you will. It would cause me to wonder what's going on with the PSA. Why is it this sensitive in making that kind of a change? And I would probably ask questions about the PSA, and what's going on, and why is it doing this? And use it as an example to ask somebody like you to explain to me why it's doing that.'' -- P4  
\end{quote}

The same judge noted one more circumstance in which the CFE might change their decision.  For each judge, we had noted that hypothetical changes might allow them to disregard risk factors that they think are not important. The judge in question responded that they generally would not use the counterfactuals to disregard risk factors. But they might use the ``sensitive'' risk factors as guides to gathering additional information (e.g., extenuating circumstances explaining a prior failure to appear) during an in-person pretrial release hearing.

\begin{quote}
``I think if [...] in a particular case I didn't think [a factor] was particularly important [I might use the counterfactual]. I mean, if it was failures to appear older than two years, but I had information in front of me that suggested, two years ago, this was the position that the defendant was in. I now know, two years later, this is the position the defendant is in, and I'm persuaded that this person has sort of turned their life around. They've been through treatment. They've kind of gotten their act together, and that maybe I therefore shouldn't treat this more than two year old stuff as seriously, then knowing that that adjustment could be made, if you take that factor out, then that adjustment could be made. I could see using it more on an individual case, as opposed to using it generally across cases, which means it really would be used, not in the context of a pretrial release decision made on paper, but a pretrial release decision made in court with the lawyers able to give me more and more meaningful information about someone's background.'' -- P4
\end{quote}

\section{Discussion} \label{sec:discussion}

\paragraph{Implications of how judges interacted with CFEs of the PSA-DMF.} 
In order to better understand whether explanations can improve human-algorithm collaborations in the context of pretrial risk assessments, we conducted think-aloud trials to study how judges use PRAIs and CFEs when making pretrial release decisions.
We hoped to learn about whether providing more insight into a PRAI's operations through CFEs, i.e., how sensitive the model is to marginal changes in defendant characteristics, would cause judges to alter their trust in the model, seek additional information, or otherwise adjust their decisions.  
For instance, one might expect reliance on a PRAI to decrease as the model's sensitivity to input factor differences increases. 

We found that CFEs did not alter the judges' decisions.
All eight judges mistakenly treated the counterfactuals as factuals rather than as indicators of model sensitivity. 
In high-stakes scenarios, there could be significant risks associated with decision-makers misinterpreting counterfactuals as factuals. For instance, if a judge detains a defendant because they misinterpret a \emph{counterfactual} prior conviction as an \emph{actual} prior conviction, the PRAI (with a CFE) would likely violate the defendant's due process rights.  
Furthermore, once judges did understand what the CFEs meant, they ignored them (rather than used them to understand the model's sensitivity).  This behavior was consistent across all eight judges interviewed, and suggests that achieving the potential benefits from CFEs will not be straightforward. 

More broadly, our findings suggest that there are many challenges associated with improving human-algorithm collaborations using explanations. In response to evidence about the errors and biases of high-stakes algorithms, many policymakers have promoted explanations to help people understand algorithmic advice and use it responsibly \cite{green2022flaws}.\footnote{We note that the term explanation can carry many meanings.  Here, we are referring to local explanations about a specific decision.  Calls for transparency about the overall workings of an algorithm before and during deployment fall into a different (and better trod) category.}
However, both laypeople and practitioners have been shown to interact with explanations in ways that diverge from what engineers and policymakers expect.
Our findings build on prior work demonstrating that, despite their intuitive appeal, algorithmic explanations do not always improve people’s ability to interpret and act on algorithmic recommendations (e.g., \citet{bansal2021does, Green2019Principles, jacobs2021machine, lai2019human}).

One important implication of our work is that explanations should be developed to align with the types of reasoning that the intended users employ in practice. 
Although judges struggled to understand the CFEs we provided, they are in fact familiar with counterfactual reasoning. 
As law students, judges are trained to think counterfactually. 
For example, first-year students are taught to reason through a negligence lawsuit in part by analyzing ``but-for'' causation. This analysis asks whether the plaintiff would still have been harmed had the defendant acted more reasonably. 
Many employment discrimination cases similarly depend on whether a manager would have made the same adverse employment decision if the plaintiff's race or gender were different. 

However, CFEs do not align with the counterfactual reasoning that judges typically employ. 
Legal counterfactuals help lawyers make direct comparisons between what a litigant did and hypothetical alternative actions. In contrast, the CFE presented changes to defendant profiles---aspects of their criminal history that were outside of the defendant's control by the time they were arrested for the present charge(s).
These explanations were intended to show how the model's output would have changed following specific edits to a profile that the defendant (by definition) could not have changed. 
Because counterfactual thinking in judicial decision-making is usually limited to alternative actions the defendant could have taken before being sued or criminally charged, judges first misinterpreted our counterfactuals as \emph{real} changes to a defendant's choices and thus their profiles.
When reminded that the counterfactuals could not have been real because defendant profiles are always fixed, the judges ignored them. 
In other words, judges are likely to dismiss alternative information about the PRAI because nothing has changed or could have changed about the defendant's past, and the release decision does not depend on counterfactual thinking about the defendant's current behavior. 

This gap between the counterfactual logic familiar to judges and counterfactual logic presented by our explanations could explain why the judges struggled to understand the CFEs and ignored the explanations once they understood them.
In order for CFEs to improve human decision-making, it will likely be necessary to align the explanations with the counterfactual logic already employed by practitioners operating in that context. 
This may also require much more comprehensive training for practitioners. 

\paragraph{Implications of how judges interacted with the PSA (without CFEs).} We also gathered information about how judges perceive/interact with the PSA-DMF without CFEs.%\footnote{As a reminder, we first presented PSA-DMF recommendations without any counterfactuals, and added the counterfactuals only after judges made an initial release decision.}
Our interviews suggest that judges view the PSA-DMF as having a relatively small influence on their decision-making process. 
Judges stated that they primarily focus on the case/charge information and the attributes of defendants, i.e., the risk factors.
They also expressed doubts about the utility of the PSA algorithm. 
These interviews align with ethnographic studies showing that judges and other criminal justice practitioners resist the use of algorithms, in part due to a sense of professional autonomy and fear of deskilling (e.g., \citet{brayne2020technologies}). 

However, these claims by judges may not tell the full story. 
First, several judges also stated that they defer to the PSA-DMF in certain circumstances;
even judges who claimed to focus on the case details said that they review the PSA scores before making a final decision and often defer to them. 
Second, prior work shows people tend to underestimate the effects that algorithmic advice has on their behavior \cite{Green2019Disparate, Green2019Principles, green2021alter}.
Thus, even if judges state that they do not closely follow the PSA-DMF, the instrument could still influence their behavior in ways that they do not recognize (e.g., via anchoring/automation biases). 

One particularly surprising comment was P3's statement that they have been asked to defer to the PSA when making decisions. 
As noted above, this statement contradicts the claim by Arnold Ventures and policymakers that judges are not required or expected to follow DMF recommendations \cite{arnold2019faqs, nj2017report}. 
This statement also does not reflect the exact instructions given at the implementation training. 
Rather, judges were told that, although they retain final discretion, if they never followed the model’s recommendations, it would be impossible to evaluate the model’s efficacy.\footnote{Direct observation from one of this paper's authors, who attended the training.} 
In other words, this judge has internalized a stronger directive to follow the DMF recommendations than the training explicitly provided. 
This (mis)interpretation complicates the claims that judges retain full discretion when making pretrial decisions. 
It is therefore necessary to evaluate not just the explicit instructions provided by pretrial risk assessment developers, but also how judges interpret and act on those instructions. 
Even if judges do not view an instruction to follow the PSA-DMF recommendation as binding, such an instruction could generate automation bias, which would diminish the role of their independent judgment in decision-making.

\paragraph{Limitations.}
Our study was small, with a study population of just eight judges. 
However, we observed a high level of consistency in how all eight judges responded to CFEs.
Given the current paucity of user studies for CFEs, the detailed reactions described in this paper provide useful insights into how experienced decision-makers respond to counterfactuals. 
Moreover, these findings can inform future work involving larger-scale user studies with CFEs.

Secondly, we explored just one approach to presenting CFEs. 
Judges' initial confusion about the meaning of CFEs might have been a function of our UX design. 
Similarly, the judges' reluctance to use the hypothetical information provided by the counterfactuals could have resulted from insufficient training at the start of the survey. 
After observing the responses of the first four judges, we attempted to address this possibility by adding an interactive presentation that provided more information about what the CFEs mean and how they could be used.
Following this training, all four Round 2 judges were able to summarize what the explanations meant.
However, this additional training did not alter the judges' decisions-making.
It is possible that more in-depth training or alternative presentations of the counterfactual information would have led to significantly different behavior.  
An important direction for future work will be to explore whether alternative methods for presenting CFEs would lead to different outcomes.

Our study was also limited to specific jurisdictions where the PSA-DMF has been in use.
We chose judges from such locations intentionally, so that our participants would already be familiar with the basic PSA-DMF before adding CFEs.
However, the judges' prior experience with the PSA-DMF could have affected how they responded to the CFEs.
Because these judges had previously used the PSA-DMF without explanations, they might have been confused about how to incorporate the CFEs into their reasoning.
We attempted to account for this potential issue by providing more instructions to participants in Round 2, which helped judges understand the CFEs during training more quickly than in Round 1.
However, once the judges encountered CFEs about specific cases and understood that they were hypothetical, they largely ignored them (as they did in Round 1). 
The fact that judges exhibited the same behaviors even after additional training suggests that there might be a fundamental conflict between CFEs and judicial reasoning, rather than just a lack of understanding of the CFEs.

Finally, it is important to recognize that judicial think-aloud responses might not provide complete insight into their behaviors in practice. 
Judges might behave differently if they were making decisions with real-world stakes.  
Moreover, what judges tell us about their reasoning may not reflect their actual reasoning. 
People are notoriously bad at accurately describing their cognitive processing \cite{nisbett1977telling}---a phenomenon that has also been observed in human collaborations with algorithms \cite{Green2019Disparate, Green2019Principles, green2021alter}. 
In the case of judges, there may be particularly strong motivations to downplay the influence of PRAIs out of fear that these tools diminish their autonomy and professional status. 
Any disconnect between how judges think they use and how they actually use algorithms could have significant consequences. If judges mistakenly believe that a PRAI does not influence their behavior, they may be less likely to scrutinize the advice that it provides.
It is therefore essential to pursue mixed-methods research on how people use algorithms and explanations, cross-referencing qualitative insights (both think-aloud and ethnographic) with quantitative analyses of human behavior. 

\paragraph{Future Work.}
Our results suggest several directions for future work. 
First and foremost, future work should evaluate CFEs with a larger sample of judges and a wider sample of decision-makers. 
Such work could investigate whether judges uniquely find CFEs unintuitive and unhelpful. 
The think-aloud results provided here should be further interrogated through large-scale experimental studies that evaluate whether and how CFEs alter human decision-making. 
Such studies should explore methods for teaching users what CFEs represent; our results suggest that even highly educated and experienced decision-makers can struggle to understand what the explanations mean. 
In order to thoroughly study how people use CFEs, we must first be able to present CFEs in ways that are comprehensible and actionable. 

Second, future work should consider alternative methods for presenting CFEs.
Our study explored just one particular design for displaying CFEs. 
As P4 suggests, decision-makers might find CFEs more helpful if they can (interactively) query a system for information about specific changes rather than simply receive information containing a long list of changes. 
For instance, if a judge could ask the model how its recommendation would change in light of a specific alteration to a defendant's profile---perhaps in light of extenuating circumstances for a prior failure to appear---judges might find the CFEs more intuitive and more helpful.

\paragraph{Conclusion.} 
We conducted think-aloud trials to study how judges use PRAIs and CFEs when making pretrial release decisions. 
We found that judges initially mistook the counterfactuals as factual changes to defendants.
Once the judges understood the CFEs, they essentially ignored them. 
This behavior was consistent across all eight judges interviewed.
Our findings suggest that using (at least these kinds of) explanations to improve human and AI collaboration is not straightforward.
These results also highlight the importance of evaluating XAI systems with their intended users. 
In the context of PRAI-assisted decision-making, we suspect the gap between the counterfactual logic judges encounter in their legal training and the logic of CFEs presented a challenge, explaining the judges' surprising yet consistent reactions.
Future work should therefore explicitly consider the types of reasoning employed by the intended users when developing XAI systems (including the instructions and UX).

\section*{Acknowledgements}
YY acknowledges support from IBM Research, the Miami Foundation, and NSF IIS-1750358. We thank Derek Bamabuer, Zana Buçinca, Isaac Lage, Todd Proebsting, and Andrew Woods for helpful feedback and discussions.

\bibliography{references}

\appendix
\counterwithin{figure}{section} % Appendix figures are referenced A.1, etc.

\section{Round 1 Survey} \label{sec:round1-survey}

In this section, we describe the Round 1 study design and think-aloud protocol.

\subsection{Defendant Data} \label{sec:defendant-data}
Our goal was to create cases that were as realistic as possible.  We obtained a sample of 500 de-identified cases from a county in Iowa that had used the PSA-DMF, consisting of actual defendants' statutory charges, risk factor values, and PSA-DMF recommendations. 
Using these defendants, we reduced the sample to a cohort of male defendants charged with non-violent crimes, for the purpose of reducing response variance. 
We then divided the cohort into three bins: one for which the corresponding CFEs included only less restrictive release recommendations, one for which they included only restrictive release recommendations, and one in which they included both more and less restrictive releases recommendations. 
Within each of these bins, we then grouped defendants by risk factors, and selected, from each of the largest groups, a set of defendants with the most common charges. 

This process ensured that the defendant profiles overlapped, but were also unique in terms of risk factors, charge information, and the number of counterfactual changes leading to more/less restrictive release recommendations. 
For each of these defendants, we synthesized demographic information, including the date of birth, date of arrest, race, and name (ensuring the name was likely perceived to correspond to the defendant's race, using the top first names identified by~\citet{gaddis2017} and top last names identified by~\citet{word2008}). Lastly, we ensured that each study consisted of two defendants identical in every way except for their race and name (pairs were not presented adjacently).

\subsection{Think-Aloud Protocol}

\paragraph{Before the study:} [The researchers introduce themselves]. Thank you so much for taking the time to chat with us today! 

Today we were hoping to have you take a survey that we designed (it has instructions so it should be completely self-explanatory) and to have you think aloud and narrate your thought process as you take the study. We are very interested in better understanding your thought process. Additionally, if you can let us know if anything doesn't make sense, if you would present information differently, or if there’s anything else, that can help us improve the study. This will be super helpful for us to ensure the study measures what we want it to measure, as well as for it to help support the quantitative results from your and others' answers (with an understanding of the process that led to them).

As you take the study, we are hoping to record the meeting about your thought process to augment our notes.  If we refer back to it, the recording will only be seen by the research team. If we end up including some of your feedback into a future publication (e.g., quotations), it will of course be anonymized. Does that sound OK to you?  [If YES, start recording]. Do you have any questions before we begin?

\paragraph{During the study:}
\begin{itemize}
    \item If participants stop thinking aloud, we prompt them to do so. 
    \item If participants repeatedly skip a section/information block, we prompt them to at least ask why they aren't paying attention to it.
    \item When the participants finish evaluating each defendant, we ask them: Did you bring in any additional information/inferences from your prior experience to inform your decisions?
\end{itemize}

\paragraph{After the study:}
Thank you again for taking the time to give us this feedback, it's super helpful. If you don't mind, we just have a few questions to ask you before we wrap up:
\begin{itemize}
    \item Were there any parts of the study you found confusing?
    \item After seeing the ``explanation'' to the model, how did your responses change?
    \item Did you find the ``explanation'' helpful?
    \item Without the think-aloud, approximately how much time would you take per defendant?
\end{itemize}

\subsection{Survey} \label{sec:round1-survey-diff}

The survey for Round 1 follows the same format as the one for Round 2 (in \cref{sec:round2-survey}), with the following exceptions:

\paragraph{Qualitative Questions.} In the Round 1, we asked participants to consider how fair the algorithmic recommendation was and how much they relied on each component of the PSA-DMF report (defendant demographic info, case charge and description, risk factors, risk scores, and release recommendation). After presenting the CFE (for the same defendant) and asking them whether this new information changed their original release decision, we additionally asked them which components of the PSA-DMF report they relied on when thinking about whether to change their original decision. 
These qualitative questions were asked on a 1-5 Likert scale. 
For Round 2, we omitted these question from the survey. 

\paragraph{Explanation of Counterfactuals.} We only provided a description of what the explanation was and omitted any instructions about how one may use it.

\section{Round 2 Survey} \label{sec:round2-survey}

In this section, we describe the think-aloud study protocol and include a full copy of the survey for Round 2 of the study in  \cref{sec:round2-survey-screenshots}. 

\subsection{Think-Aloud Protocol}

\paragraph{Before the study:} [The researchers introduce themselves]. Thank you so much for taking the time to chat with us today! 

Today, we were hoping that you will take a survey that we designed and have you think aloud and narrate your thought process as you take the study. We know it can be a little strange to narrate your thoughts as you go, but we are specifically interested in learning more about your thought process. Hearing your thought process will also help us ensure that the study measures what we want it to measure, as well as support the quantitative results from your and others' answers.

As you take the study, we were hoping to record this meeting via Zoom. The recording will only be seen by the research team, if at all. If we end up including some of your feedback into a future publication (e.g., quotations), it will of course be anonymized. Does that sound OK to you?  [If YES, start recording].

Do you have any questions before we begin? Feel free to interrupt us if anything I say doesn't make sense! Next, we will go over the instructions for the survey. After every section of the instructions, we'll pause and just ask you to explain back what we said---this is not to test or quiz you, just to make sure we're explaining this well enough and that we're on the same page, because some of what follows can be a tad tricky. 

\paragraph{Instructions:} [Display a sample PSA-DMF report]. In this study, you will be given information about an arrestee, as well as the output of the Public Safety Assessment (PSA). Based on this information, you will then be asked whether you think the arrestee should be released pending case disposition, and if so under what conditions. Here's the part of the PSA report that displays the arrestee's demographic information, information about the current charge, and the risk factors. Take a moment to familiarize yourself with the format. And here's the model's risk scores and recommendation---again, take a moment to familiarize yourself with the format.

[Display CFE]. In addition to the arrestee's information and the PSA report, you will also be given a set of hypothetical changes to the arrestee's profile and associated changes to the model recommendations. These hypothetical changes create counterfactuals: different versions of the same arrestee's that aren't true. This information can help you understand how much each factor affects the model's recommendation, which can help you decide whether you would stick with your initial release decision about the arrestee or change it. 

Specifically, the model explanation section gives you intuition for whether the model's recommendation changes a lot or very little with one hypothetical change to the arrestee's profile. The left side depicts scenarios in which a hypothetical change to the arrestee's criminal history would cause the recommendation's severity to decrease, and the right side depicts scenarios in which a hypothetical change to the arrestee's criminal history would cause the recommendation's severity to increase. 

For example, the arrestee whose report you just saw is (1) 23 years old or older, and (2) has two prior failures to appear within the last two years. For this arrestee, the model originally recommended release option D (with face-to-face contact every two weeks, court reminder notifications, etc.). The box at the very bottom and to the left shows a hypothetical situation in which that very same arrestee had instead failed to appear only once before and how the model's recommendation would significantly drop in severity to release option B (with only one phone contact per month and court reminders). Alternatively (at the very bottom and to the right), if the same arrestee were younger (even 21 or 22 years old), the model's recommendation would increase in severity to release option E, which includes electronic monitoring, curfews, and other forms of supervisions. 

So let me ask you a few questions about what we've shown you so far---this is not to quiz you, just to make sure that we've explained everything well enough so far:
\begin{itemize}
    \item What are the bottom left and right boxes displaying?
    \item If this arrestee hypothetically had another pending charge, what would the model recommend?
\end{itemize}

So, to summarize again, these boxes show whether the model's overall recommendations are sensitive to changing a single feature of the arrestee's profile, keeping everything else the same. It might also lead you to depart from the model's recommendation depending on how important a factor is to you in your decision-making. 

Now, let me ask you a few questions about what we've shown you so far---this is again not to quiz you, just to make sure that we've explained everything well enough so far:
\begin{itemize}
    \item If this arrestee is bordering on 23 years of age at the time of arrest, what are the possible recommendations the model would produce? 
    \item For this arrestee, if you decide ``failures to appear pretrial in past 2 years'' should not weigh heavily in your decision (e.g., because of some extenuating circumstances), would you depart from the model's recommendation?
\end{itemize}
We just want to note that, for this particular arrestee, there are hypotheticals leading to both less restrictive and more restrictive release decisions, on the left and right, respectively. Not all arrestees will have hypotheticals in which a single change to their profile causes the model to make a different recommendation.
Do you have any questions before we continue?

\paragraph{During the study:}
\begin{itemize}
    \item If the participants stop thinking aloud, we prompt them to do so. 
    \item If the participants repeatedly skip a section/information block, we prompt them to at least ask why they aren't paying attention to it.
    \item At the end of every arrestee profile, we ask the participants: did you bring in any additional information/inferences from your prior experience to inform your decisions?
\end{itemize}

\paragraph{After the study:}
Thank you again for taking the time to give us this feedback, it's super helpful. If you don't mind, we just have a few questions to ask you before we wrap up:
\begin{itemize}
    \item Were there any parts of the study you found confusing?
    \item After seeing the ``explanation'' to the model, how did your responses change? 
    \item Did you find the ``explanation'' helpful? 
    \item Lastly, without the think-aloud, approximately how much time would you take per arrestee?
\end{itemize}

\section{Correctness on Comprehension Questions} \label{sec:correctness-comprehension}

\begin{figure*}[p]
    \centering
    \includegraphics[width=0.96\textwidth]{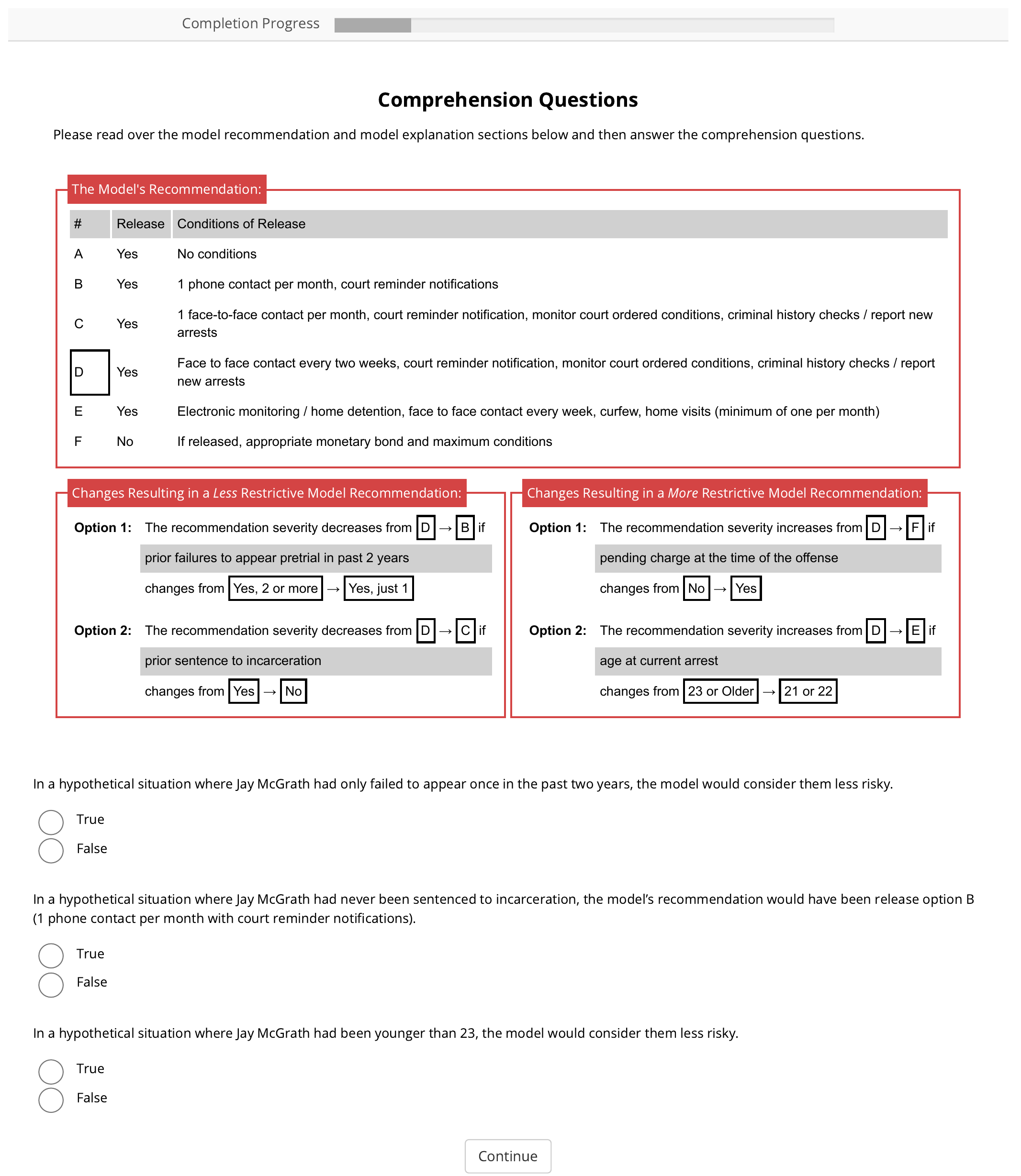} 
    \caption{\textbf{Round 2 Survey -- Comprehension Questions.} Note that all defendant and booking information in this study is synthesized.}
    \label{fig:r2-comprehension1}
\end{figure*}

We composed three comprehension questions, designed to ensure that the participants understood the information presented to them in the CFEs.
The questions can be found in \cref{fig:r2-comprehension1}.
All participants were able to complete the first and third of these questions easily after one attempt, and half answered the second question correctly on the first attempt (with the other half answering correctly after a second try).
From our think-aloud sessions, we have no reason to suspect that participants who initially missed the second question did not understand the information presented in the CFE.
Rather, we merely believe the question was a bit tricky.
The question asks whether, if the defendant had never been sentenced to incarceration, the model's recommendation would change from a D to a B (the correct answer is no, since it would change from a D to a C).
All judges knew that without the prior sentence to incarceration, the model's recommendation would less restrictive, but they misread the information from the option above the one they were supposed to read, thinking it would change from a D to a B instead.
As we describe in \cref{sec:findings-explanation}, participants generally understood how to read the information in the counterfactual; they just misinterpreted the information as revisions to the defendant's profile.

\section{Screenshots from Round 2 Survey} \label{sec:round2-survey-screenshots}

\noindent\begin{minipage}{\textwidth}
    \centering
    \includegraphics[width=0.96\textwidth]{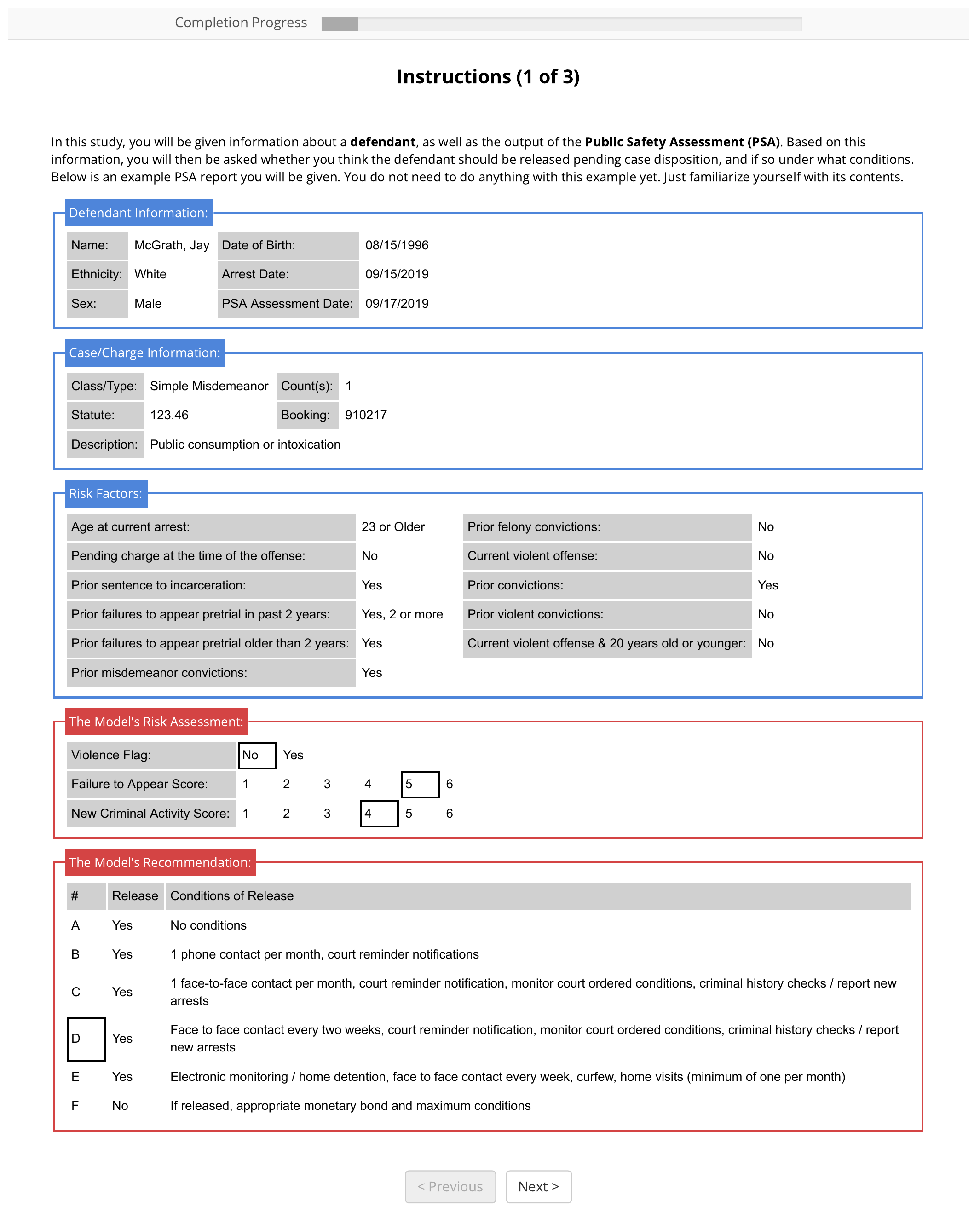} 
    \captionof{figure}{\textbf{Round 2 Survey -- Instructions 1 of 3.} The PSA report. Note that all defendant and booking information in this study is synthesized.}
    \label{fig:r2-instructions1}
\end{minipage}

\begin{figure*}[p]
    \centering
    \includegraphics[width=0.96\textwidth]{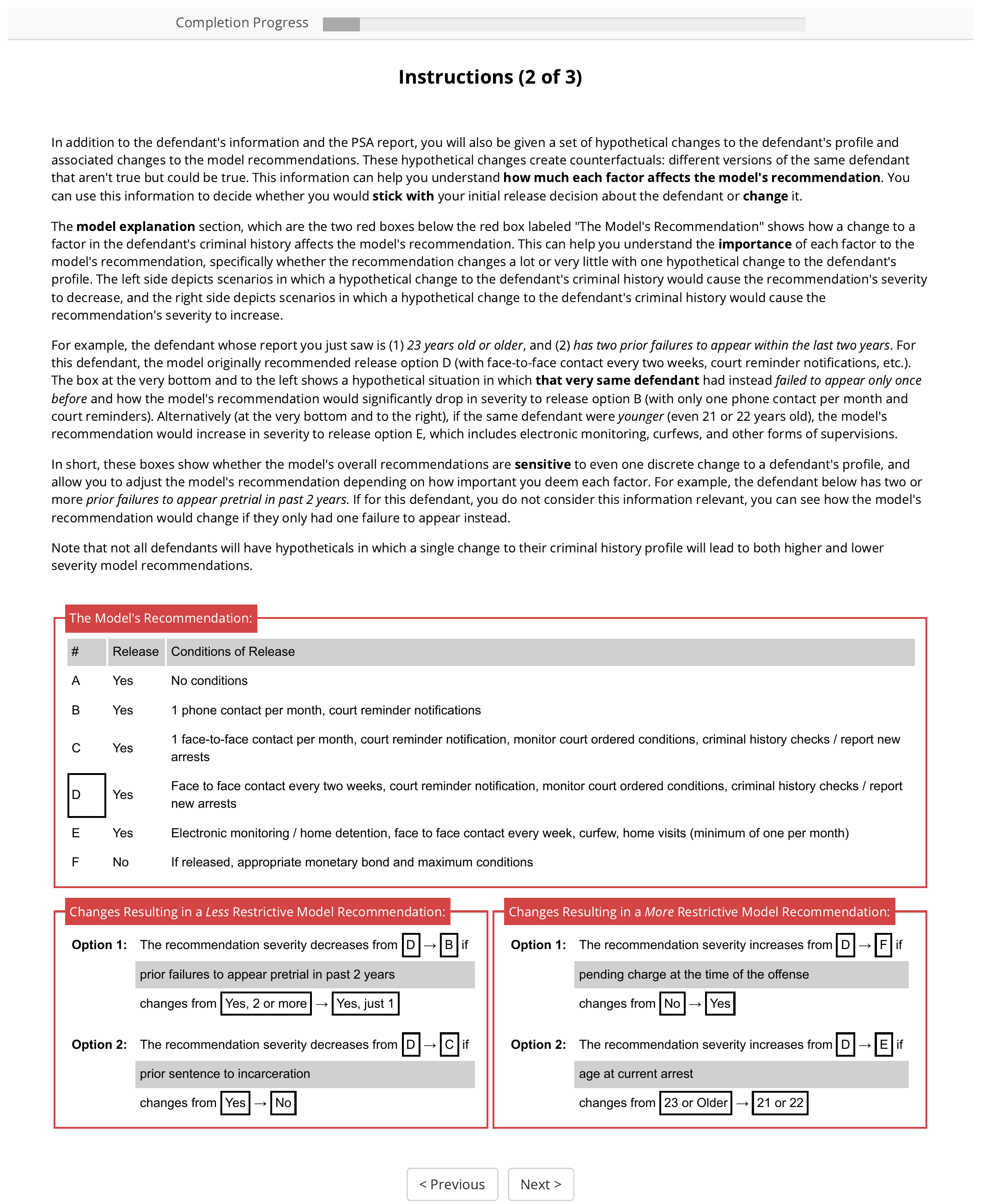} 
    \caption{\textbf{Round 2 Survey -- Instructions 2 of 3.} The counterfactual explanation.}
    \label{fig:r2-instructions2}
\end{figure*}

\begin{figure*}[p]
    \centering
    \includegraphics[width=0.96\textwidth]{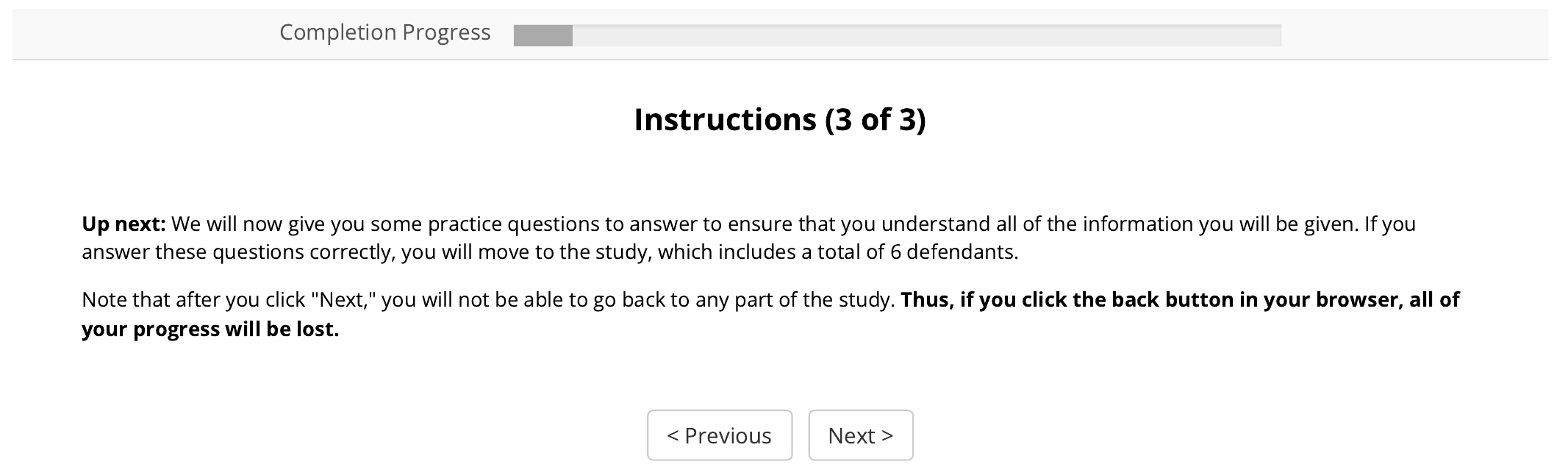} 
    \caption{\textbf{Round 2 Survey -- Instructions 3 of 3.}}
    \label{fig:r2-instructions3}
\end{figure*}

\begin{figure*}[p]
    \centering
    \includegraphics[width=0.96\textwidth]{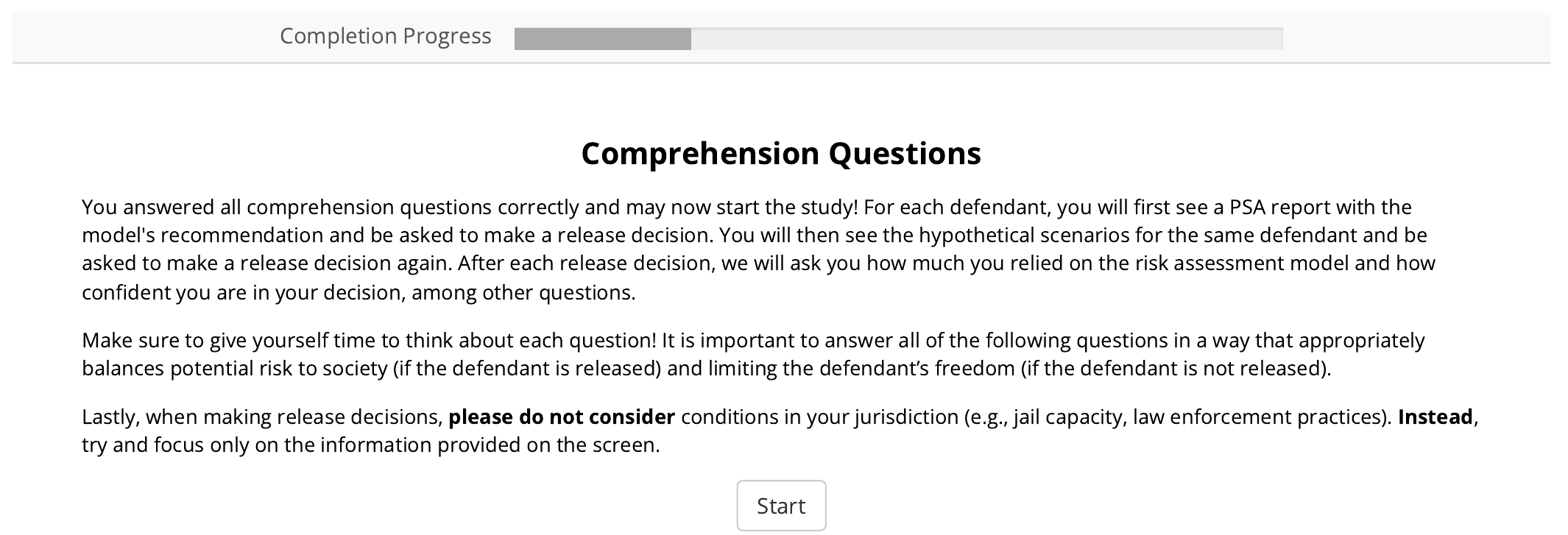} 
    \caption{\textbf{Round 2 Survey -- Instructions prior to main survey questions.}}
    \label{fig:r2-comprehension2}
\end{figure*}

\begin{figure*}[p]
    \centering
    \includegraphics[width=0.96\textwidth]{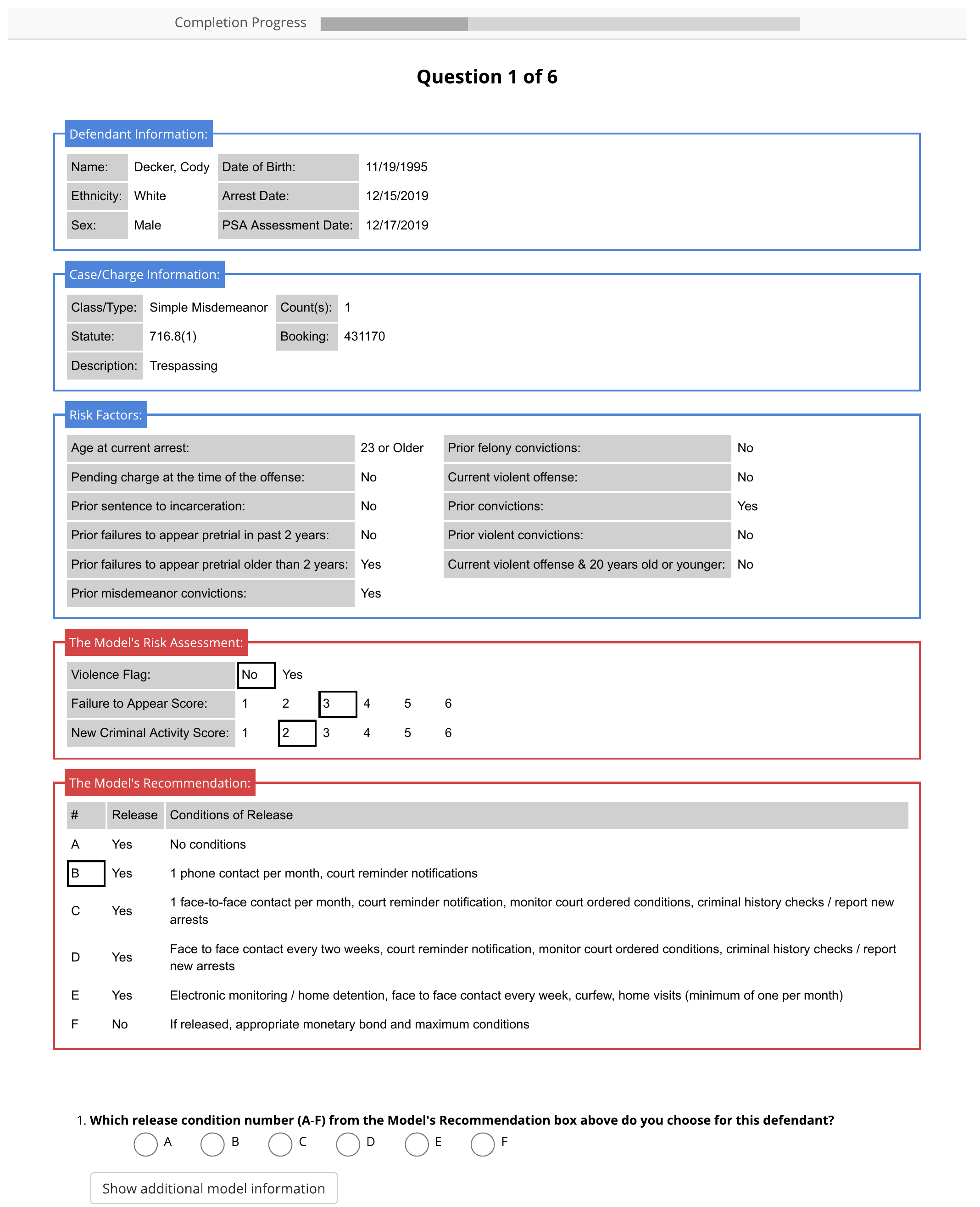} 
    \caption{\textbf{Round 2 Survey -- Question 1.} Clicking the button will reveal the counterfactual explanation. Note that all defendant and booking information in this study is synthesized.}
    \label{fig:r2-question1pre}
\end{figure*}

\begin{figure*}[p]
    \centering
    \includegraphics[width=0.96\textwidth]{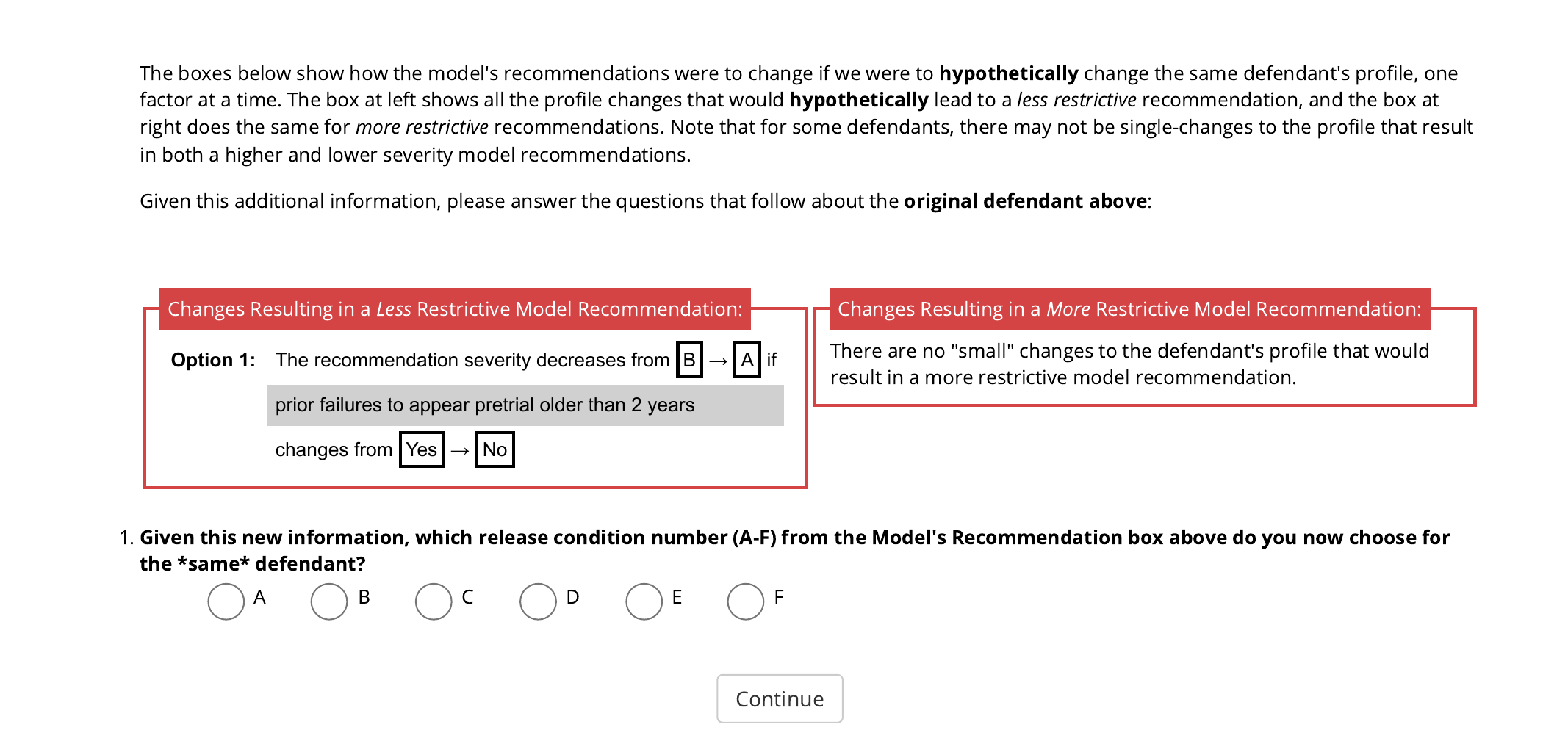} 
    \caption{\textbf{Round 2 Survey -- Question 1's counterfactual explanation.}}
    \label{fig:r2-question1post}
\end{figure*}

\begin{figure*}[p]
    \centering
    \includegraphics[width=0.96\textwidth]{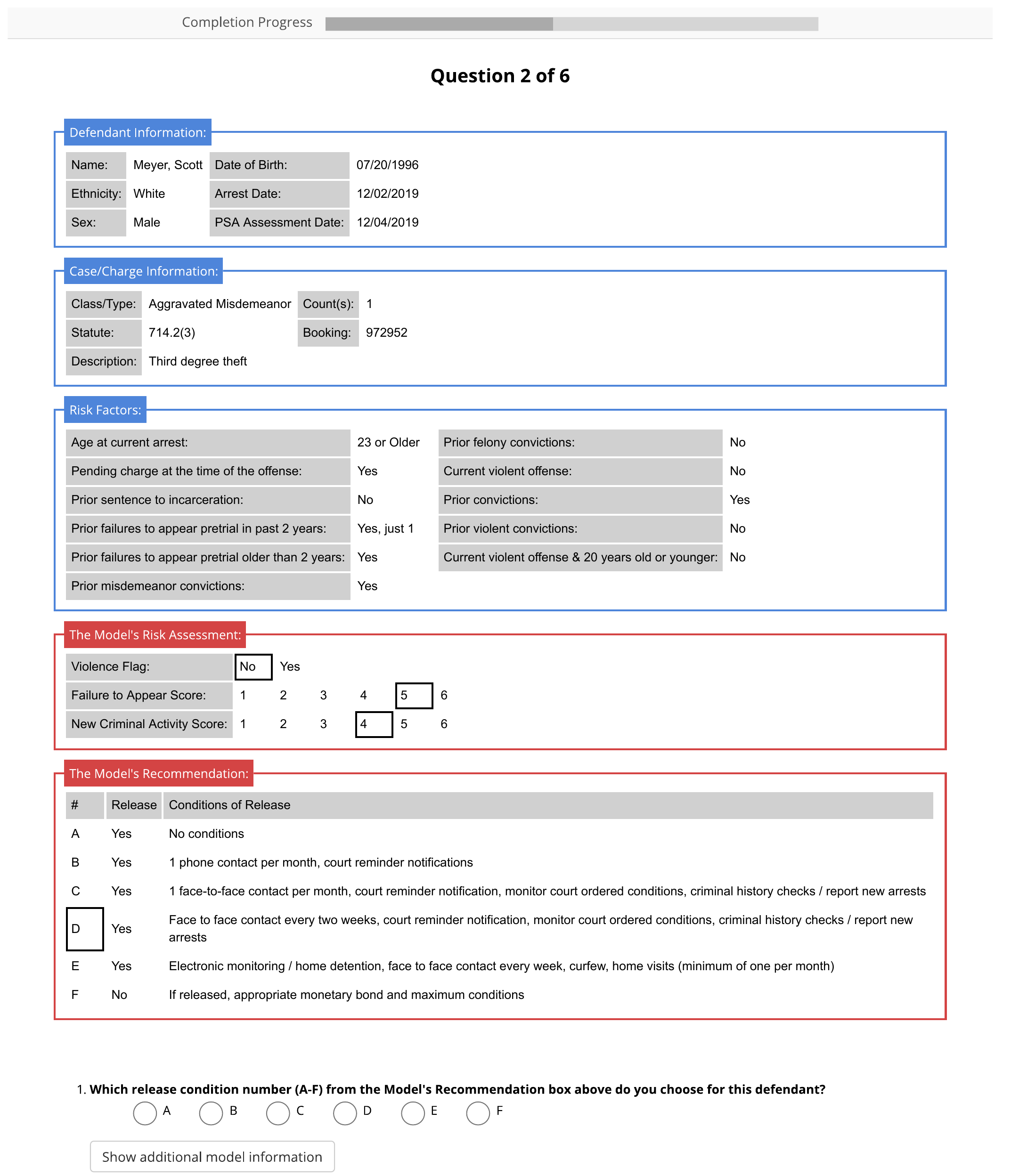} 
    \caption{\textbf{Round 2 Survey -- Question 2.} Clicking the button will reveal the counterfactual explanation. Note that all defendant and booking information in this study is synthesized.}
    \label{fig:r2-question2pre}
\end{figure*}

\begin{figure*}[p]
    \centering
    \includegraphics[width=0.96\textwidth]{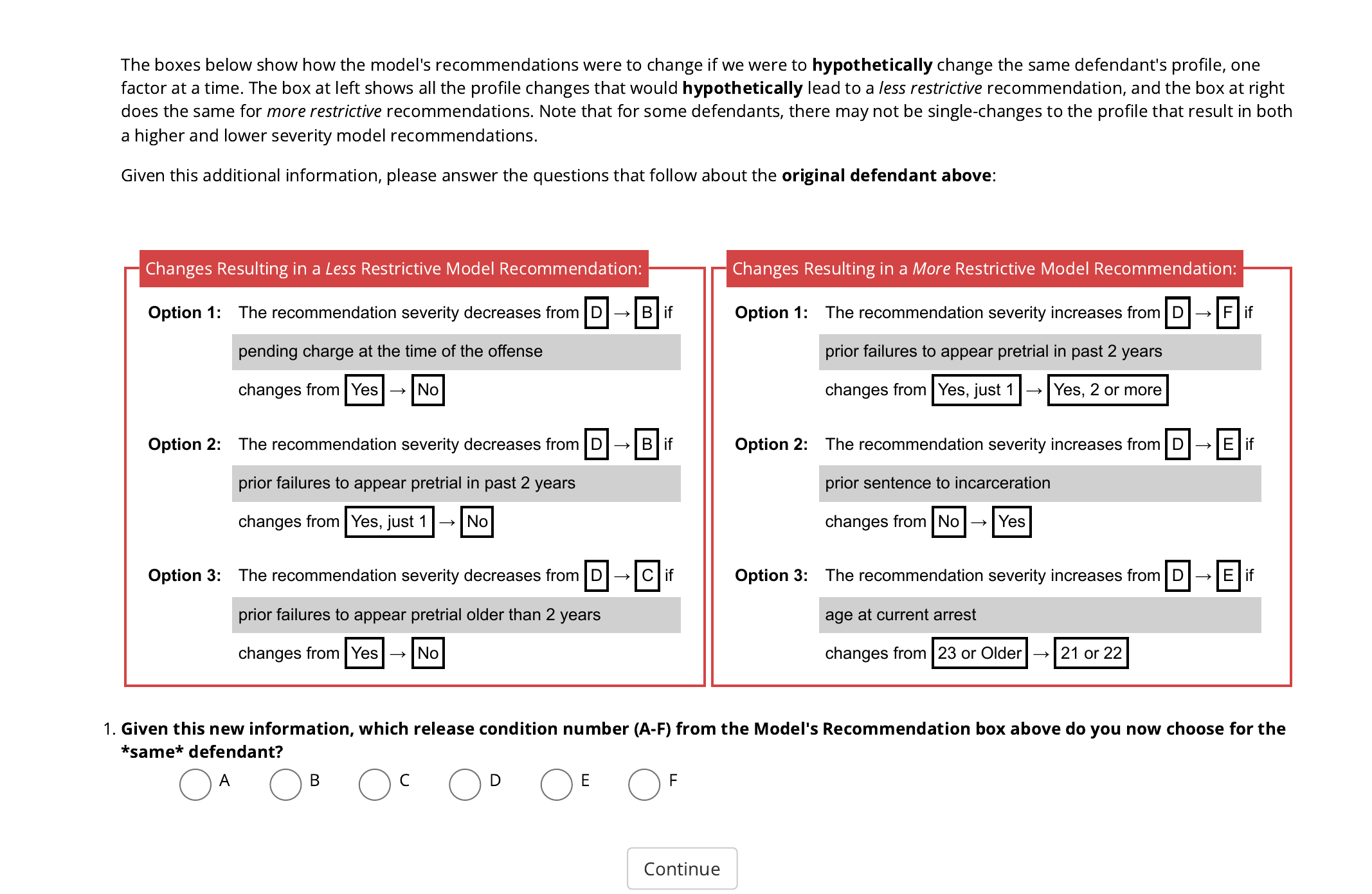} 
    \caption{\textbf{Round 2 Survey -- Question 2's counterfactual explanation.}}
    \label{fig:r2-question2post}
\end{figure*}

\FloatBarrier
\section{Additional Example Cases and Counterfactual Explanations} \label{sec:additional-cases}

\noindent\begin{minipage}{\textwidth}
    \centering
    \includegraphics[width=0.96\textwidth]{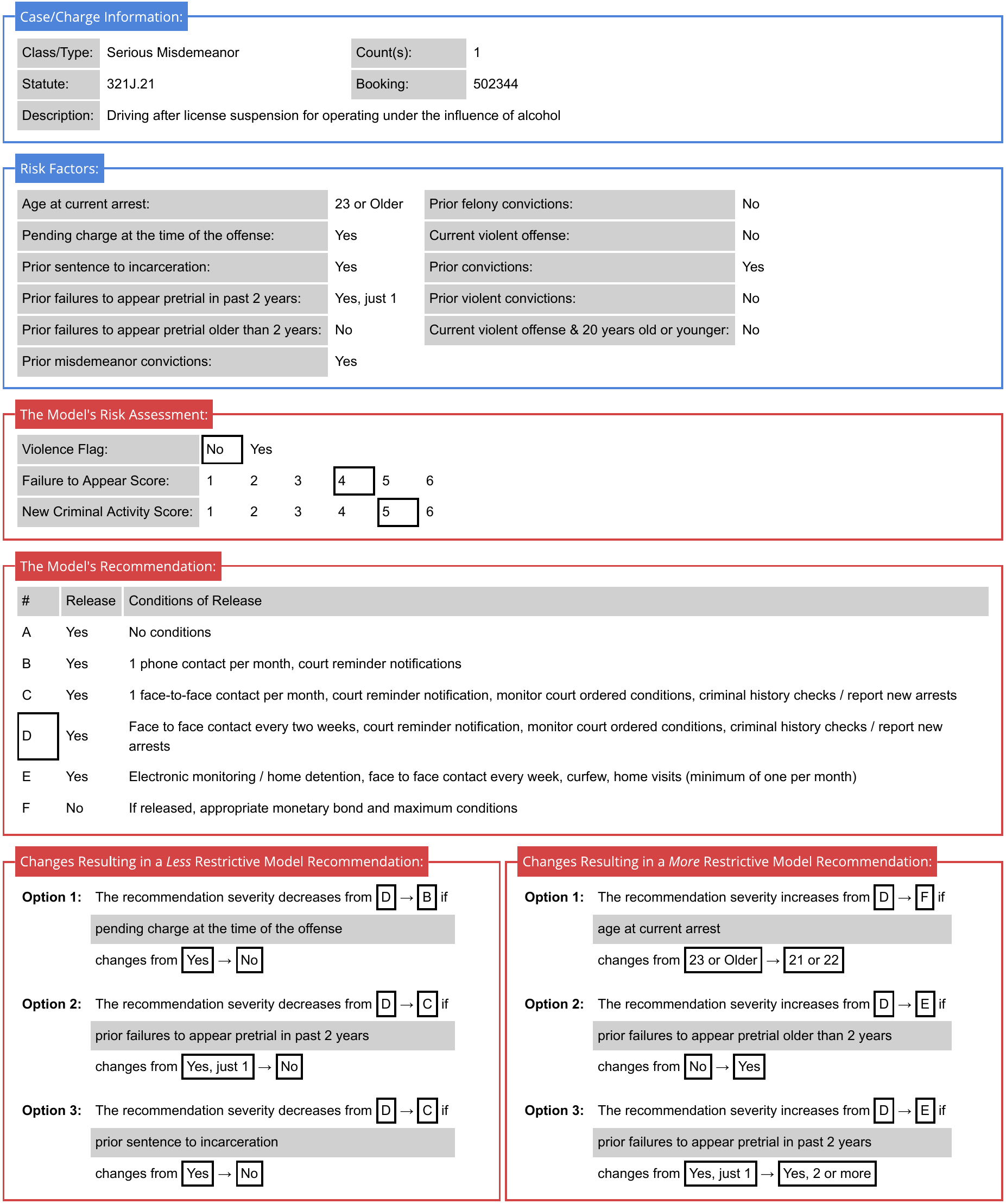} 
    \captionof{figure}{\textbf{Example PSA-DMF report and counterfactual explanation (without demographic info).}}
    \label{fig:additional-case1}
\end{minipage}

\begin{figure*}[p]
    \centering
    \includegraphics[width=0.96\textwidth]{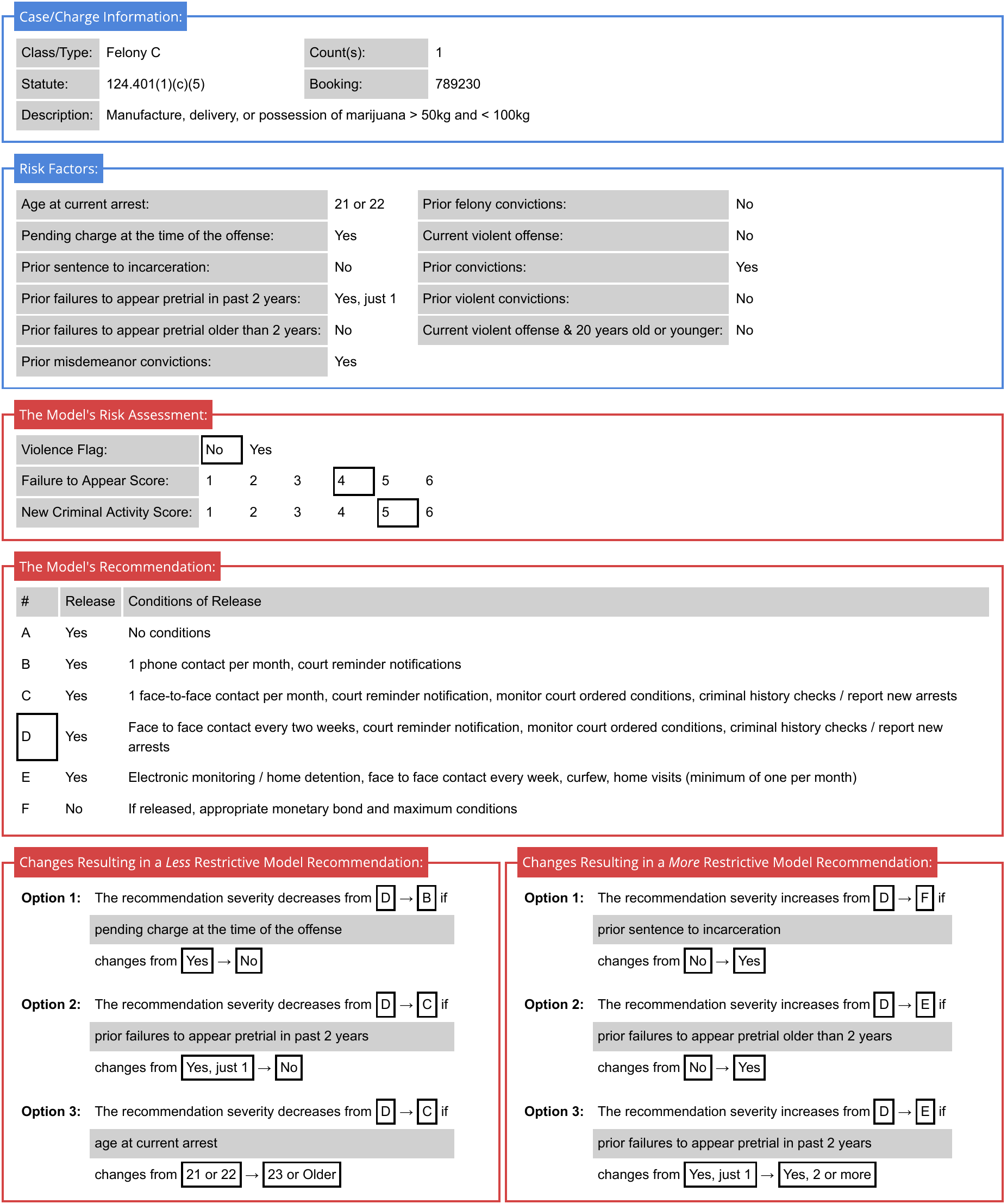} 
    \caption{\textbf{Example PSA-DMF report and counterfactual explanation (without demographic info).}}
    \label{fig:additional-case2}
\end{figure*}

\begin{figure*}[p]
    \centering
    \includegraphics[width=0.96\textwidth]{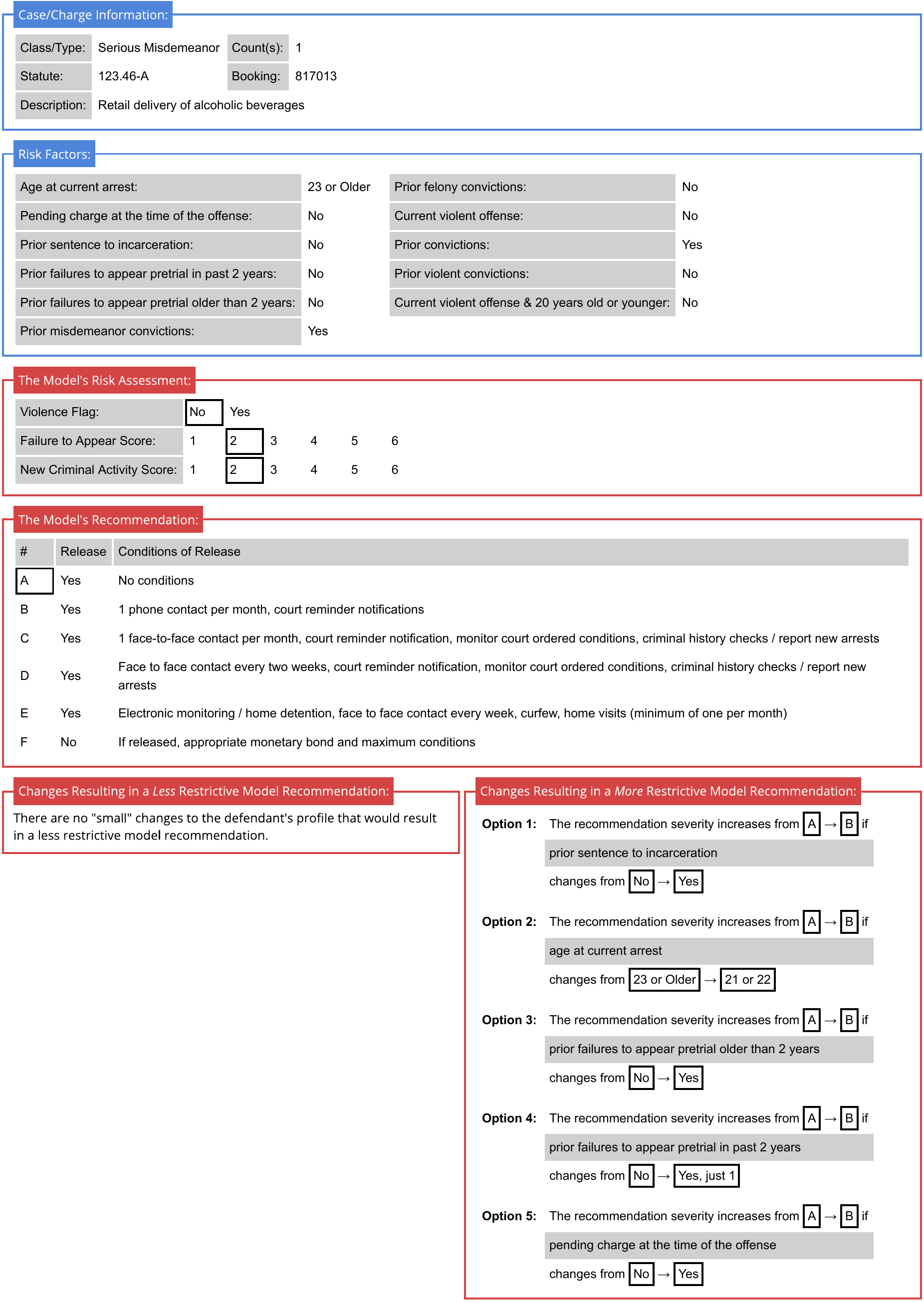} 
    \caption{\textbf{Example PSA-DMF report and counterfactual explanation (without demographic info).}}
    \label{fig:additional-case3}
\end{figure*}

\begin{figure*}[p]
    \centering
    \includegraphics[width=0.96\textwidth]{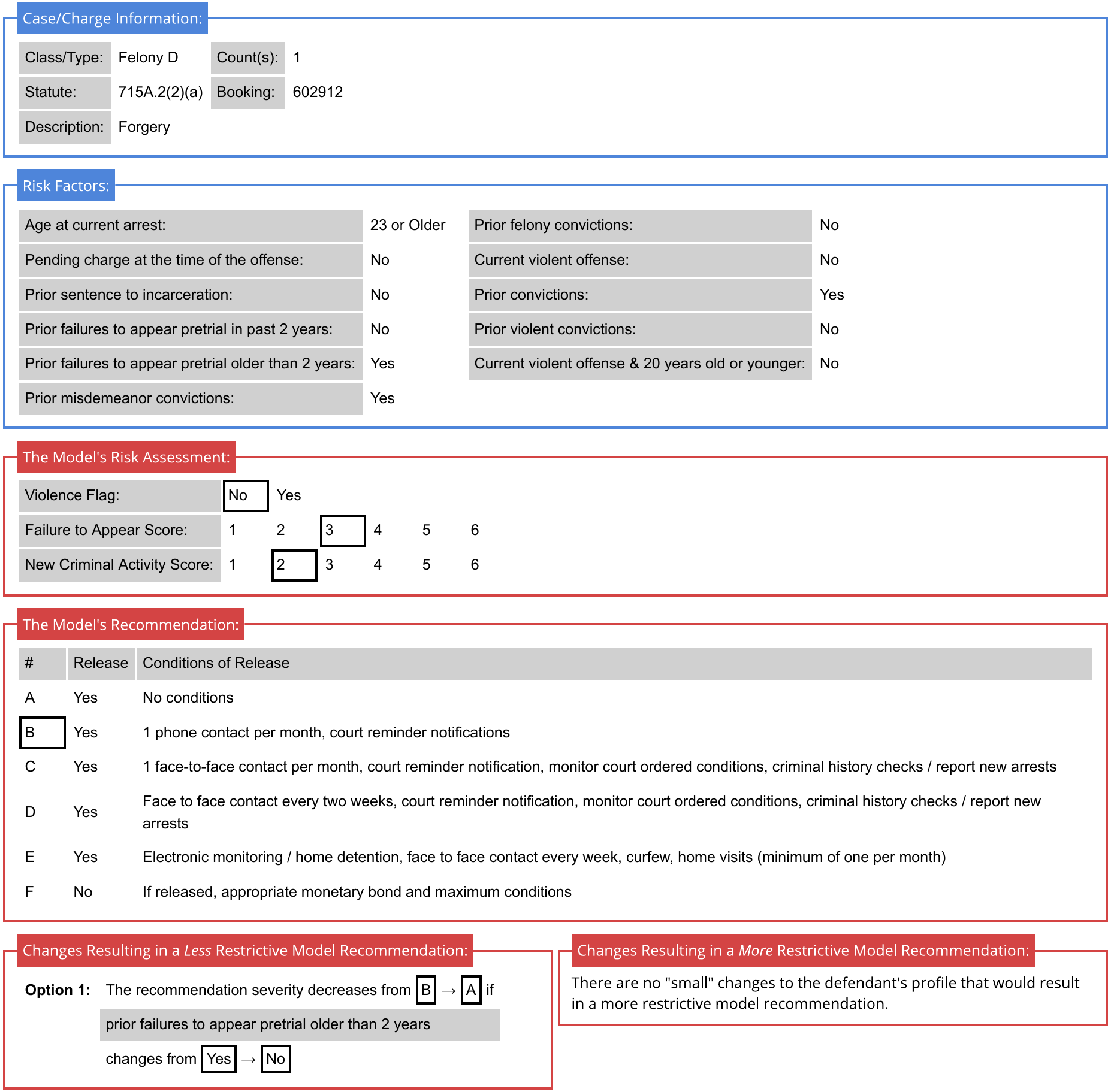} 
    \caption{\textbf{Example PSA-DMF report and counterfactual explanation (without demographic info).}}
    \label{fig:additional-case4}
\end{figure*}

\begin{figure*}[p]
    \centering
    \includegraphics[width=0.96\textwidth]{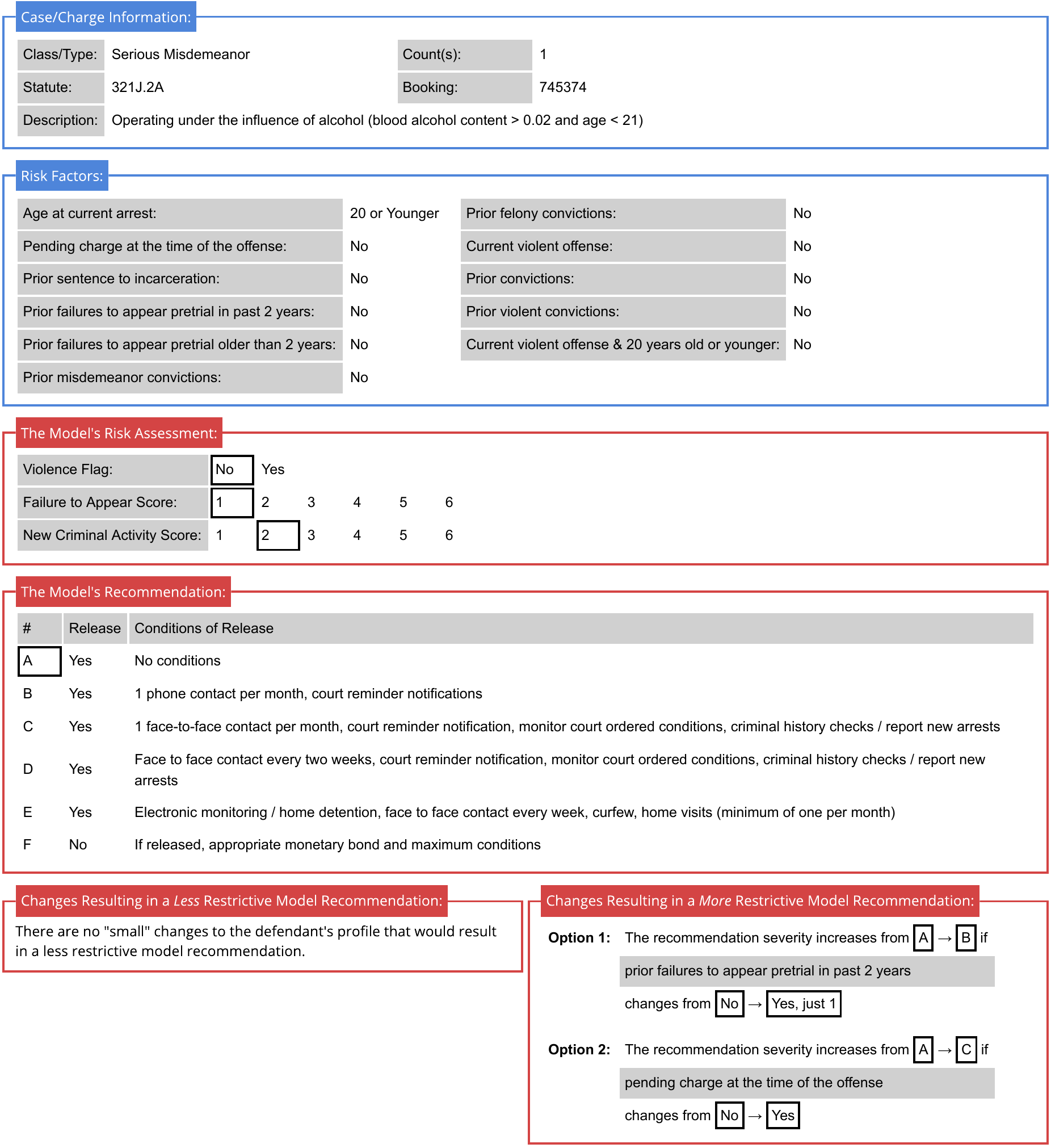} 
    \caption{\textbf{Example PSA-DMF report and counterfactual explanation (without demographic info).}}
    \label{fig:additional-case5}
\end{figure*}

\begin{figure*}[p]
    \centering
    \includegraphics[width=0.96\textwidth]{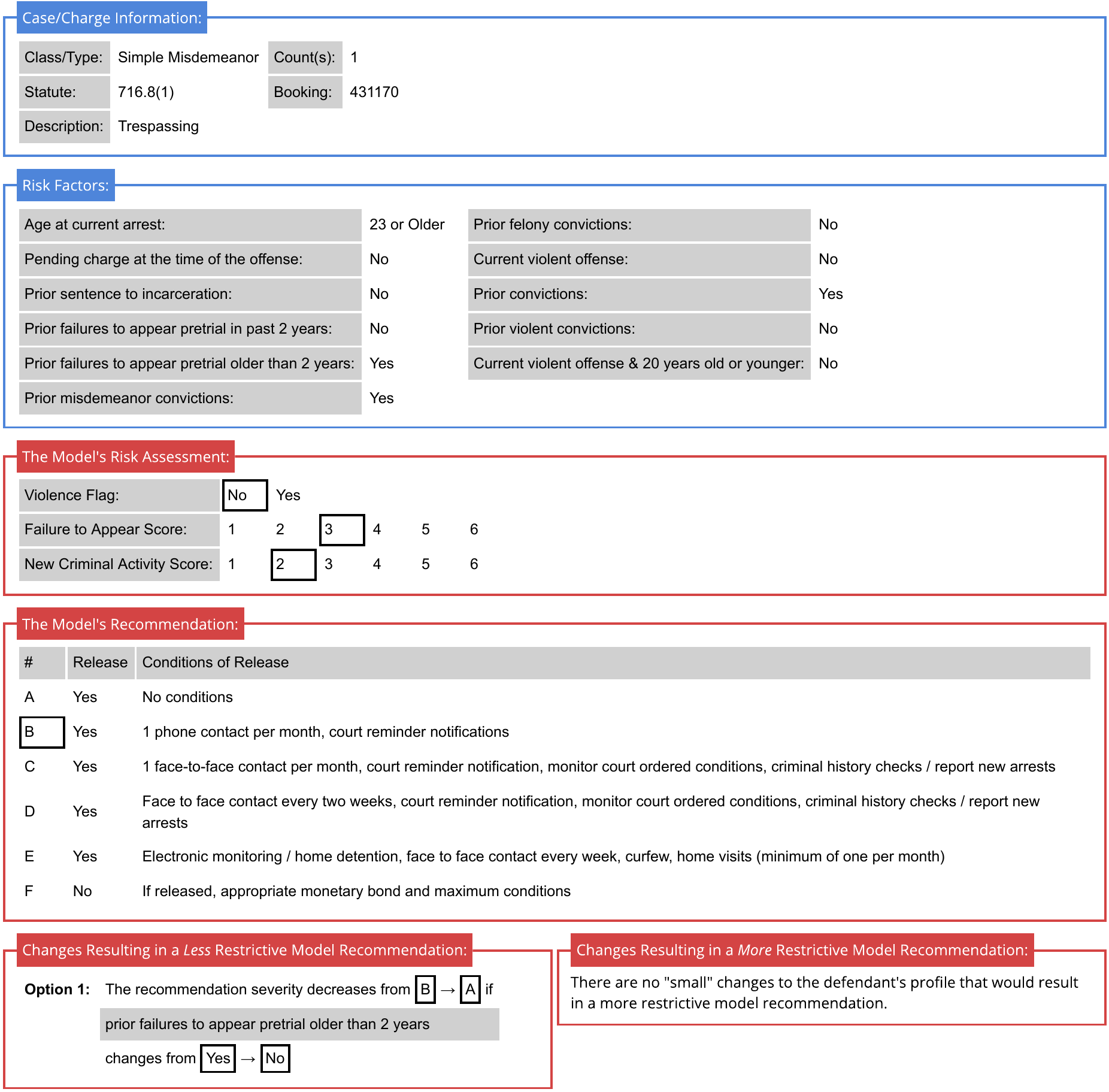} 
    \caption{\textbf{Example PSA-DMF report and counterfactual explanation (without demographic info).}}
    \label{fig:additional-case6}
\end{figure*}

\begin{figure*}[p]
    \centering
    \includegraphics[width=0.96\textwidth]{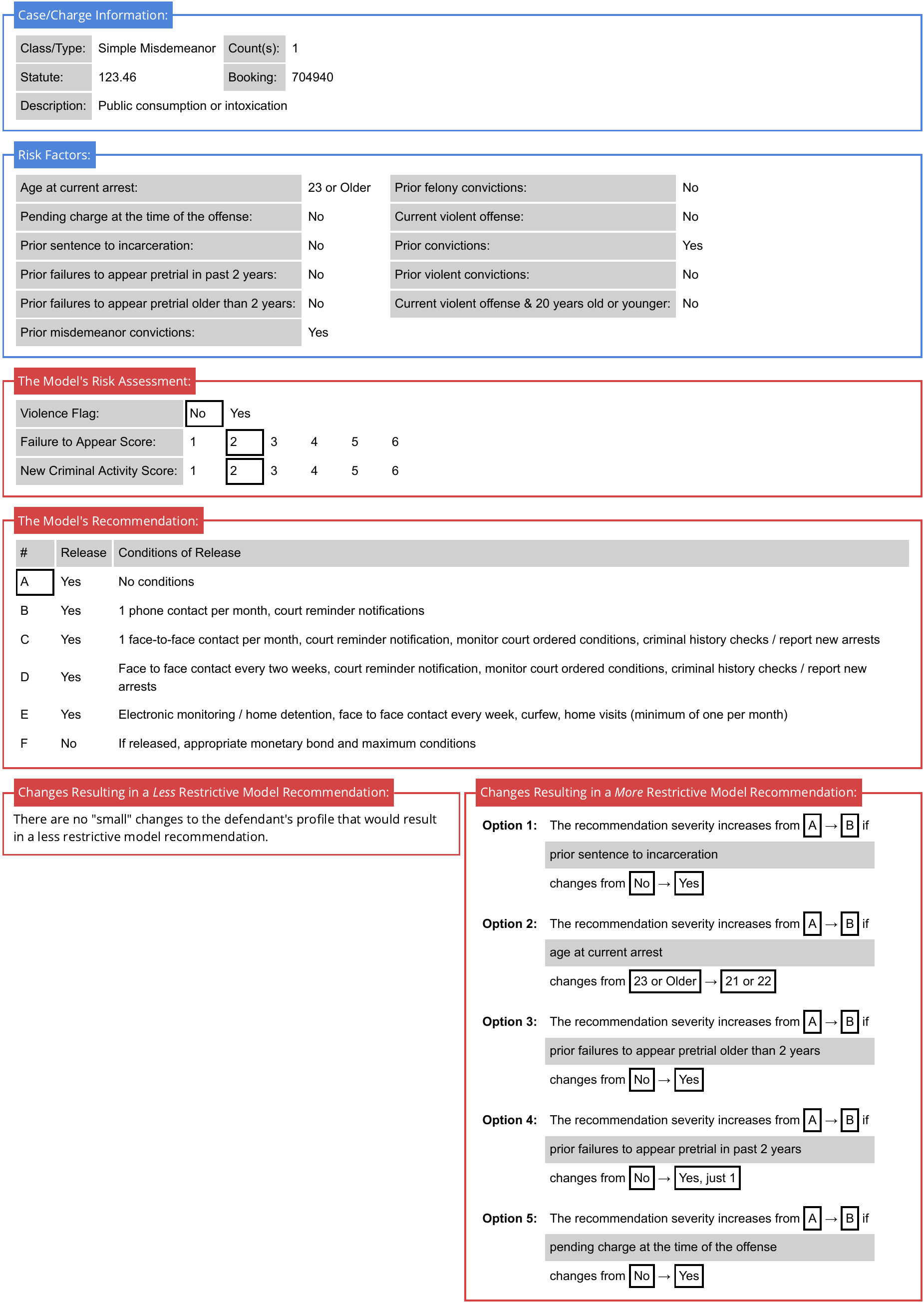} 
    \caption{\textbf{Example PSA-DMF report and counterfactual explanation (without demographic info).}}
    \label{fig:additional-case7}
\end{figure*}

\begin{figure*}[p]
    \centering
    \includegraphics[width=0.96\textwidth]{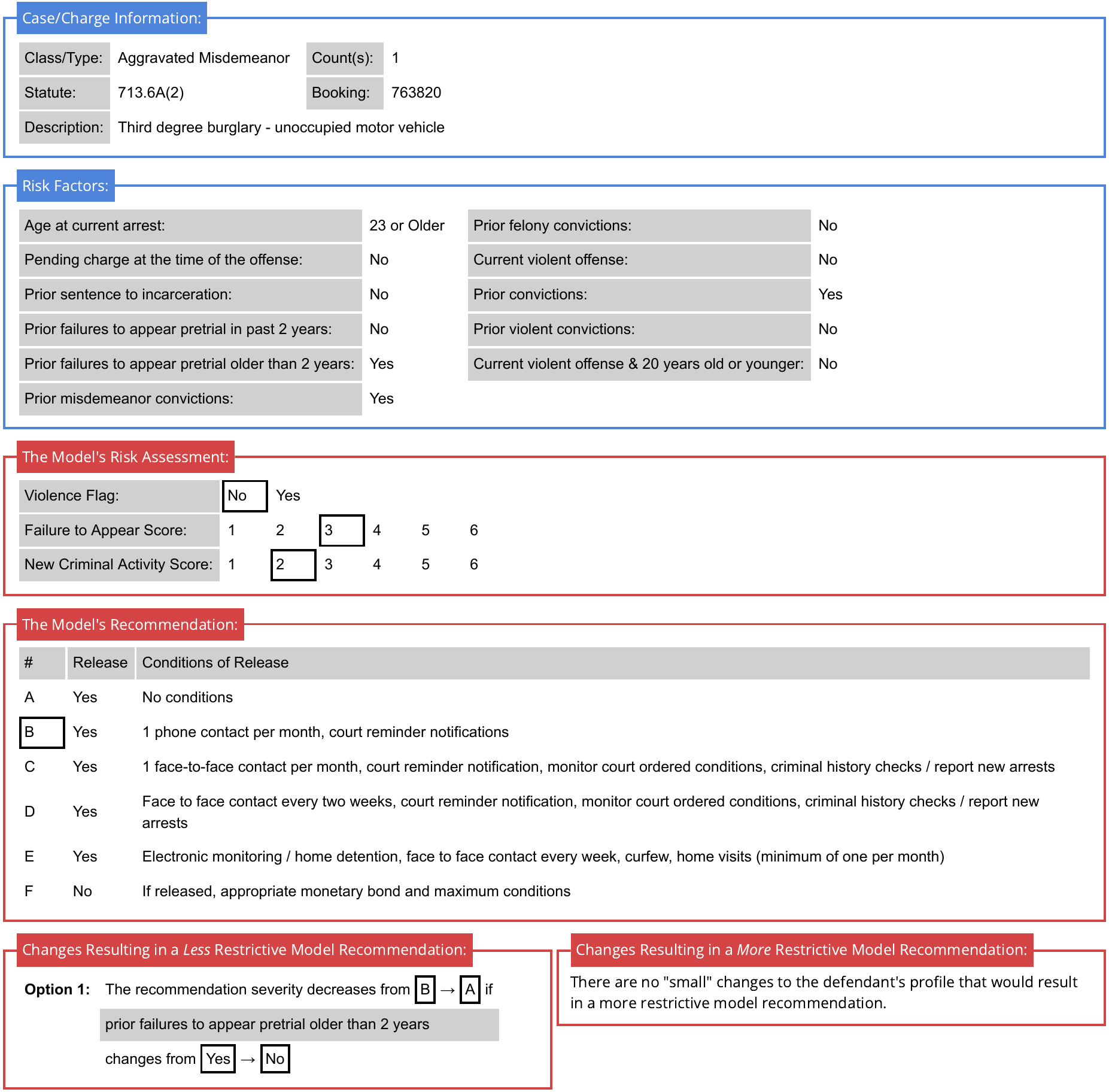} 
    \caption{\textbf{Example PSA-DMF report and counterfactual explanation (without demographic info).}}
    \label{fig:additional-case8}
\end{figure*}

\begin{figure*}[p]
    \centering
    \includegraphics[width=0.96\textwidth]{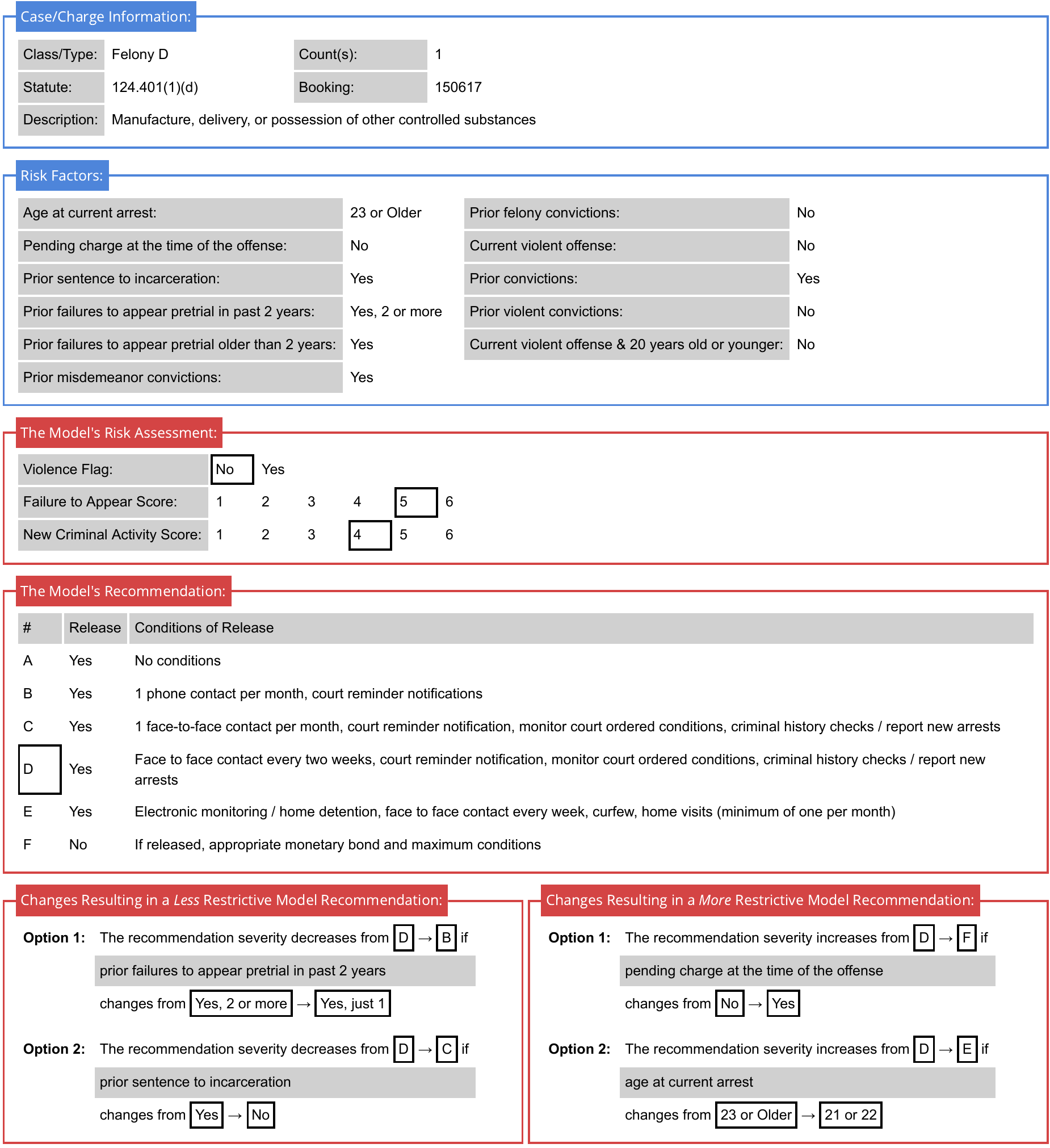} 
    \caption{\textbf{Example PSA-DMF report and counterfactual explanation (without demographic info).}}
    \label{fig:additional-case9}
\end{figure*}

\begin{figure*}[p]
    \centering
    \includegraphics[width=0.96\textwidth]{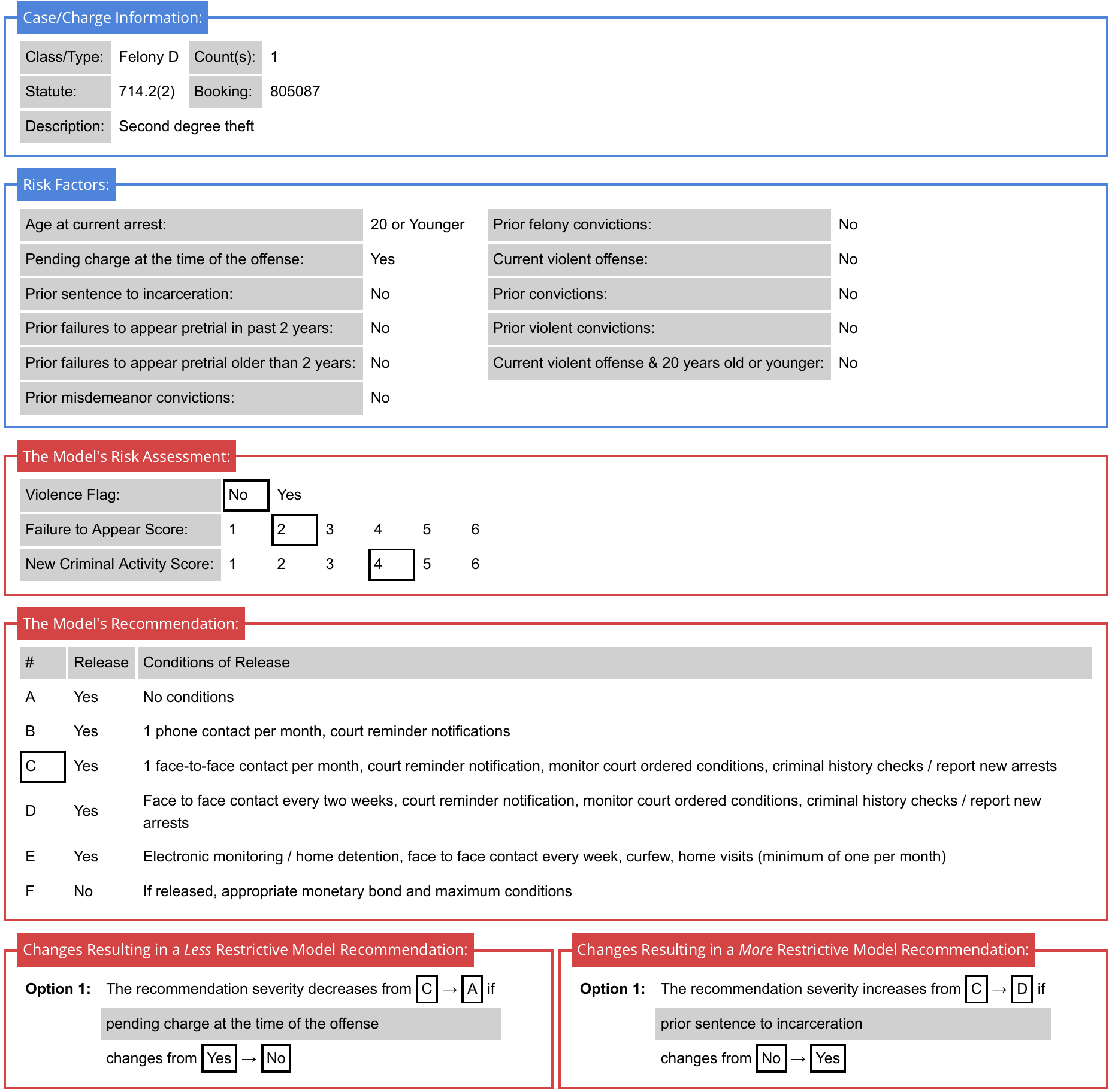} 
    \caption{\textbf{Example PSA-DMF report and counterfactual explanation (without demographic info).}}
    \label{fig:additional-case10}
\end{figure*}

\begin{figure*}[p]
    \centering
    \includegraphics[width=0.96\textwidth]{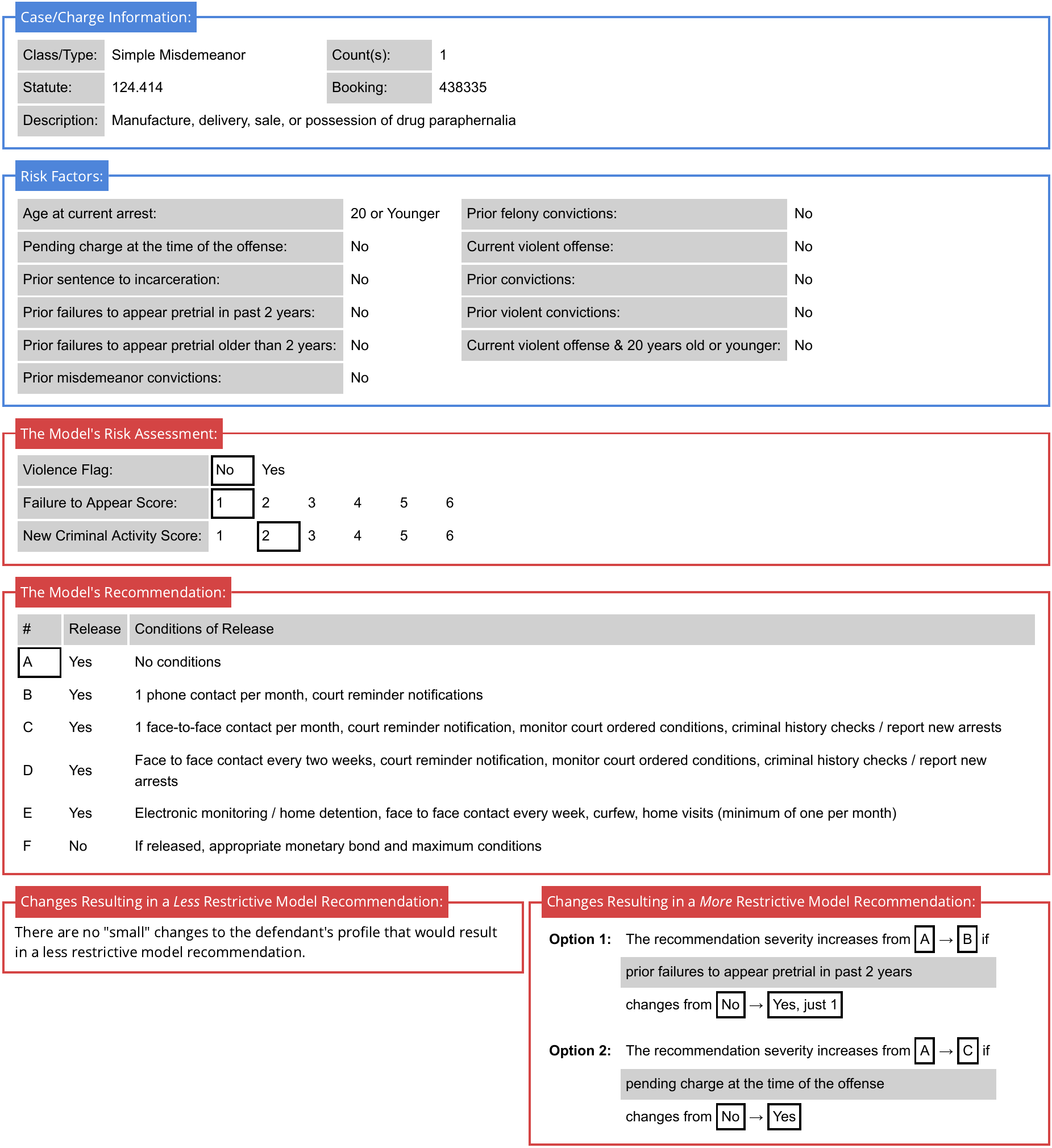} 
    \caption{\textbf{Example PSA-DMF report and counterfactual explanation (without demographic info).}}
    \label{fig:additional-case11}
\end{figure*}

\end{document}